\documentclass{article}
\usepackage{graphicx}
\usepackage{amsmath}
\usepackage{amsfonts}
\usepackage{amssymb}
\newtheorem{theorem}{Theorem}

\newtheorem{corollary}[theorem]{Corollary}

\newtheorem{definition}[theorem]{Definition}

\newtheorem{proposition}[theorem]{Proposition}

\newenvironment{proof}[1][Proof]{\textbf{#1.} }{\ \rule{0.5em}{0.5em}}

\begin{document}

\title{Wigner Particle Theory and Local Quantum Physics }
\author{Lucio Fassarella and Bert Schroer\thanks{work supported by CNPq}\\CBPF, Rua Dr. Xavier Sigaud, 150, 22290-180 Rio de Janeiro - RJ, Brazil\\email Fassarel@cbpf.br,\thinspace\thinspace\ Schroer@cbpf.br}
\date{December 2001}
\maketitle
\begin{abstract}
Wigner's irreducible positive energy representations of the Poincar\'{e} group
are often used to give additional justifications for the Lagrangian
quantization formalism of standard QFT. Here we study another more recent
aspect. We explain in this paper modular concepts by which we are able to
construct the local operator algebras for all standard positive energy
representations directly i.e. without going through field coordinatizations.
In this way the artificial emphasis on Lagrangian field coordinates is avoided
from the very beginning. These new concepts allow to treat also those cases of
\ ``exceptional'' Wigner representations associated with anyons and the famous
Wigner ``spin tower''which have remained inaccessible to Lagrangian
quantization. Together with the d=1+1 factorizing models (whose modular
construction has been studied previously), they form an interesting family of
theories with a rich vacuum-polarization structure (but no on shell real
particle creation) to which the modular methods can be applied for their
explicit construction. We explain and illustrate the algebraic strategy of
this construction.

We also comment on possibilities of formulating the Wigner theory in a setting
of a noncommutative spacetime substrate. This is potentially interesting in
connection with recent unitarity- and Lorentz invariance- preserving results
of the special nonlocality caused by this kind of noncommutativity.
\end{abstract}

\section{\bigskip The setting of the problem}

The algebraic framework of local quantum physics shares with the standard
textbook approach to QFT the same physical principles but differs in concepts
and tools used for their implementation. Whereas the standard approach is
based on ``field-coordinatizations'' in terms of pointlike fields (without
which the canonical- or functional integral- quantization is hardly
conceivable), the algebraic framework permits to formulate local quantum
physics directly in terms of a net of local operator algebras i.e. without the
intervention of the rather singular pointlike field coordinates whose
indiscriminate use is the potential source of ultraviolet divergencies. Among
the many advantages is the fact that the somewhat artistic\footnote{The
postulated canonical or functional representation requirement is known to get
lost in the course of the calculations and the physical (renormalized) result
only satisfies the more general causality/locality properties.} standard
scheme is replaced by a conceptually better balanced setting.

The advantages of such an approach \cite{Haag}\cite{Bu}\cite{Bu-Ha} were in
the eyes of many particle physicist offset by its constructive weaknesses of
which even its protagonists (who used it mainly for structural investigations
as TCP, Spin\&Statistics and alike) were well aware \cite{Bu-Ha}. In
particular even those formulations of renormalized perturbation theory which
were closest in spirit to the algebraic approach namely the causal
perturbation theory and its recent refinements \cite{Due-Fre} uses a
coordinatization of algebras in terms of fields at some stage. The underlying
``Bogoliubov-axiomatics'' \cite{Tod} in terms of an off-shell generating
``S-matrix'' S(g) suffers apparently from the same ultraviolet limitations as
any other pointlike field formulation.

However there are signs of change which are not only a consequence of the lack
of promised success of many popular attempts in post standard model particle
theory. Rather it is also becoming slowly but steadily clear that the times of
constructive nonperturbative weakness of the algebraic approach (AQFT) are
passing and the significant conceptual investments are beginning to bear
fruits for the actual construction of models.

The constructive aspects of these gains are presently most clearly visible in
situations in which there is no real (on-shell) particle creation but for
which, different from free field theories, the vacuum-polarization structure
remains very rich. It is not possible in those models to locally generate
one-particle states from the vacuum without accompanying vacuum-polarization
clouds. Besides the well-known d=1+1 factorizing models, this includes the
QFTs associated with exceptional Wigner representations i.e. d=1+2 ``anyonic''
spin and the d=1+3 ``spin towers'' (Wigner's famous exceptional zero mass
representations with an infinite number of interlinked helicity states). In
both cases the absence of compact localization renders the theories more
noncommutative and in turn less accessible to Lagrangian quantization methods.
The main content of this paper deals with constructive aspects of such models.

The historical roots of the algebraic approach date back to the 1939 famous
Wigner paper \cite{Wig} whose aim was to obtain an intrinsic conceptual
understanding of particles avoiding the ambiguous wave equation method and the
closely related Lagrangian quantization so that a physical equivalence of
different Lagrangian descriptions could be easily recognized. In fact it was
precisely this fundamental intrinsic appeal and the unicity of Wigner's
approach that some authors felt compelled to present this theory as a kind of
additional partial justification for the the Lagrangian (canonical- or
functional-) quantization \cite{Wei}. Since the late 50s there has been a
dream about a royal path into nonperturbative particle physics which starts
from Wigner's representation-theoretic particle setting and introduces
interactions in a maximally intrinsic and invariant way i.e. by using concepts
which avoid doing computations in terms of the standard singular field
coordinationations and lean instead on the unitary and crossing symmetric
scattering operator and the associated spaces of formfactors. It is well-known
that this dream in its original form failed, and that some of the old ideas
were re-processed and entered string theory via Veneziano's dual model. In the
following we will show that certain aspects of that old folklore (which
certainly does not include that of a ``Theory of Everything''), if enriched
with new concepts, can have successful applications for the above mentioned
class of models.

According to Wigner, particles should be described by irreducible positive
energy representation of the Poincar\'{e} group. In fact they are the
indecomposable building blocks of those multi-localized asymptotically stable
objects in terms of which each state can be interpreted and measured in
counter-coincidence arrangements in the large time limit. This raises the
question what localization properties particles should be expected to have,
and which positive energy representations permit what kind of localization.

There are two localization concepts. One is the ``Born-localization'' taken
over from Schroedinger theory which is based on probabilities and associated
projectors projecting onto compactly supported subspaces of spatially
localized wave functions at a fixed time (which in the relativistic context
also bears the name ``Newton-Wigner'' localization). The incompatibility of
this localization with relativistic covariance and Einstein causality was
already noted and analyzed by its protagonists \cite{N-W}. Covariance as well
as macro-causality are however satisfied in the asymptotic region and
therefore the covariance and the cluster separability of the Moeller operators
and the S-matrix are not effected by the use of this less than perfect quantum
mechanical localization. On the other hand there exists a fully relativitic
covariant localization which is intimately related to the characteristic
causality- and vacuum polarization- properties of QFT; in the standard
formulation of QFT it is that localization which is encoded in the position of
the dense subspace obtained by applying smeared fields (with a fixed test
function support) to the vacuum. Since in the field-free formulation of local
quantum physics this localization turns out to be inexorably linked to the
Tomita-Takesaki modular theory of operator algebras, it will be shortly
referred to as ``modular localization''. Its physical content is less obvious
and its consequences are less intuitive and therefore we will take some care
in its presentation.

In fact the remaining part of this introductory section is used to contrast
the Newton-Wigner localization with the modular localization. This facilitates
the understanding of both concepts.

The use of Wigner's group theory based particle concept for the formulation of
what has been called\footnote{This name was chosen in \cite{Coester} in order
to distinguish it from the field-mediated interactions of standard QFT.}
``direct interactions'' in relativistic mutiparticle systems can be nicely
illustrated by briefly recalling the arguments which led to this relativistic
form of macro-causal quantum mechanics. Bakamjian and Thomas \cite{BT}
observed as far back as 1953 that it is possible to introduce an interaction
into the tensor product space describing two Wigner particles by keeping the
additive form of the total momentum $\vec{P}$, its canonical conjugate
$\vec{X}$ and the total angular momentum $\vec{J}$ and by implementing
interactions through an additive change of the invariant free mass operator
$M_{0}$ by an interaction $v$ (with only a dependence on the relative c.m.
coordinates $\vec{p}_{rel}$) which then leads to a modification of the
2-particle Hamiltonian $H$ with a resulting change of the boost $\vec{K}$
according to
\begin{align}
M  &  =M_{0}+v,\,\,M_{0}=2\sqrt{\vec{p}_{rel}^{2}+m^{2}}\\
H  &  =\sqrt{\vec{P}^{2}+M^{2}}\nonumber\\
\vec{K}  &  =\frac{1}{2}(H\vec{X}+\vec{X}H)-\vec{J}\times\vec{P}%
(M+H)^{-1}\nonumber
\end{align}
The commutation relations of the Poincar\'{e} generators are maintained,
provided the interaction operator $v$ commutes with $\vec{P},\vec{X}$ and
$\vec{J}.$ For short range interactions the validity of the time-dependent
scattering theory is easily established and the Moeller operators $\Omega
_{\pm}(H,H_{0})$ and the $S$-matrix $S(H,H_{0})$ are Poincar\'{e} invariant in
the sense of independence on the L-frame
\begin{equation}
O(H,H_{0})=O(M,M_{0}),\,\,O=\Omega_{\pm},S
\end{equation}
and they also fulfill the cluster separability%
\begin{equation}
s-\lim_{\delta\rightarrow\infty}O(H,H_{0})T(\delta)\rightarrow\mathbf{1}%
\end{equation}
where the $T$ operation applied to a 2-particle vector separates the particle
by an additional spatial distance $\delta.$ The subtle differences to the
non-relativistic case begin to show up for 3 particles \cite{Coester}. Rather
than adding the two-particle interactions one has to first form the mass
operators of the e.g. 1-2 pair with particle 3 as a spectator and define the
1-2 pair-interaction operator in the 3-particle system
\begin{align}
M(12,3)  &  =\left(  \left(  \sqrt{M(12)^{2}+\vec{p}_{12}^{2}}+\sqrt
{m^{2}+\vec{p}_{3}^{2}}\right)  ^{2}-\left(  \vec{p}_{12}+\vec{p}_{3}\right)
^{2}\right)  ^{\frac{1}{2}}\\
V^{(3)}(12)  &  \equiv M(12,3)-M(1,2,3),\,\ M(1,2,3)\equiv M_{0}(123)\nonumber
\end{align}
where the notation speaks for itself (the additive operators carry a subscript
labeling and the superscript in the interaction $V^{(3)}(12)$ operators remind
us that the interaction of the two particles within a 3-particle system is not
identical to the original two-particle $v\equiv V^{(2)}(12)$ operator in the
two-particle system). Defining in this way $V^{(3)}(ij)$ for all pairs, the
3-particle mass operator and the corresponding Hamiltonian are given by
\begin{align}
M(123)  &  =M_{0}(123)+\sum_{i<j}V_{{}}^{(3)}(ij)\\
H(123)  &  =\sqrt{M(123)^{2}+p_{123}^{2}}\nonumber
\end{align}
and lead to a L-invariant and cluster-separable 3-particle Moeller operator
and S-matrix, where the latter property is expressed as a strong operator
limit
\begin{align}
S(123)  &  \equiv S(H(123),H_{0}(123))=S(M(123),M_{0}(123))\\
&  s\text{-}\lim_{\delta\rightarrow\infty}S(123)T(\delta_{13},\delta
_{23})=S(12)\times\mathbf{1}\nonumber
\end{align}
with the formulae for other clusterings being obvious. By iteration and the
use of the framework of rearrangement collision theory (which introduces an
auxiliary Hilbert space of bound fragments), this can be generalized to
n-particles including bound states \cite{C-P}.

As in nonrelativistic scattering theory, there are many different relativistic
direct particle interactions which lead to the same S-matrix. As Sokolov
showed, this freedom to modify off-shell operators (e.g. $H,\vec{K}$ as
functions of the single particle variables $\vec{p}_{i},\vec{x}_{i},\vec
{j}_{i}$ and the interaction $v$) may be used to construct to each system of
the above kind a ``scattering-equivalent'' system in which the
interaction-dependent generators $H,\vec{K}$ restricted to the images of the
fragment spaces become the sum of cluster Hamiltonians (or boosts) with
interactions between clusters being switched off \cite{C-P}. Using these
interaction-dependent equivalence transformations, the cluster separability
can be made manifest. It is also possible to couple channels in order to
describe particle creation, but this channel coupling ``by hand'' does not
define a natural mechanism for interaction-induced vacuum polarization.

Even though such direct interaction models between relativistic particles can
hardly have fundamental significance, their very existence as relativistic
theories (i.e. consistent with the physically indispensible macro-causality)
help us rethink the position of micro-causal and local\ versus nonlocal but
still macro-causal relativistic theories.

Since our intuition on theses matters is notoriously unreliable and ridden by
prejudices, it is very useful to have such illustrations. This is of
particular interest in connection with recent attempts to implement
nonlocality through noncommutativity of the spacetime substrate (see the last
section). But even some old piece of QFT folklore, which claimed that the
construction of unitary relativistic invariant and cluster-separable
S-matrices can only be achieved through local QFT, are rendered incorrect.

It turns out that if one adds crossing symmetry to the list of S-matrix
properties it is possible to show that if the on-shell S-matrix originates at
all from a local QFT, it determines its local system of operator algebras
uniquely \cite{S-matrix}. This unicity of local algebras is of course the only
kind of uniqueness which one can expect since individual fields are analogous
to coordinates in differential geometry (in the sense that passing to another
locally related field cannot change the S-matrix).

The new concept which implements the desired crossing\ property and also
insures the principle of ``nuclear democracy``\footnote{Every particle may be
interpreted as bound of all others whose fused charge is the same. An explicit
illustration is furnished by the bootstrap properties of d=1+1 factorizing
S-matrices \cite{Kar}.} (both properties are not compatible with the above
relativistic QM) is modular localization. In contrast to the quantum
mechanical Newton Wigner localization, it is not based on projection operators
(which project on quantum mechanical subspaces of wave functions with support
properties) but rather is reflected in the Einstein causal behavior of
expectation values of local variables in modular localized state vectors.
Modular localization in fact relates off-shell causality, interaction-induced
vacuum polarization and on-shell crossing in an inexorable manner and in
particular furnishes the appropriate setting for causal propagation properties
(see next section). Since it allows to give a completely intrinsic definition
of interactions in terms of the vacuum polarization clouds which accompany
locally generated one-particle states without reference to field coordinates
or Lagrangians, one expects that it serves as a constructive tool for
nonperturbative investigations. This is borne out for those models considered
in this paper.

It is interesting to note that both localizations are preempted in the Wigner
theory. Used in the Bakajian-Thomas-Coester spirit of QM of relativistic
particles with the Newton-Wigner localization, it leads to relativistic
invariant scattering operators which obey cluster separability properties and
hence are in perfect harmony with macro-causality. On the other hand used as a
starting point of modular localization one can directly pass to the system of
local operator algebras and relate the notion of interaction (and exceptional
statistics) inexorably with micro-causality and vacuum polarization clouds
which accompany the local creation of one particle states. Perhaps the
conceptually most surprising fact is the totally different nature of the local
algebras from quantum mechanical algebras.

In the second section we will present the modular localization structure of
the standard halfinteger spin Wigner representation in the first subsection
and that of the exceptional (anyonic, spin towers) representations in the
second subsection.

The subject of the third section is the functorial construction of the local
operator algebras associated with the modular subspaces of the standard Wigner
representations. The vacuum polarization aspects of localized particle
creation operators associated with exceptional Wigner representations are
treated in the fourth section. In section 5 we explain our strategy for the
construction of theories which have no real particle creation but (different
from free fields) come with a rich vacuum polarization structure in the
context of d=1+1 factorizing models.

Apart from the issue of anyons, the most interesting and unexplored case of
QFTs related to positive energy Wigner representations is certainly that of
the massless d=1+3 ``Wigner spin towers''. This case is in several aspects
reminiscent of structures of string theory. It naturally combines all (even,
odd, supersymmetric) helicities into one indecomposable object. If it would be
possible to introduce interactions into this tower structure, then the
standard argument that any consistent interacting object which contains spin 2
must also contain an (at least a quasiclassical) Einstein-Hilbert action
(which is used by string theorist in order to link strings with gravity)
applies as well here \footnote{In this connection it appears somewhat ironic
that the infinite spin tower Wigner representation is often dismissed as ``not
used by nature'' without having investigated its physical potential.}.

Recently there has been some interest in the problem whether the Wigner
particle structure can be consistent with a noncommutative structure of
spacetime where the minimal consistency is the validity of macro-causality. We
will have some comments in the last section.

\section{Modular aspects of positive energy Wigner representations}

In this in the next subsection we will briefly sketch how one obtains the
interaction-free local operator algebras directly from the Wigner particle
theory without passing through pointlike fields. The first step is to show
that there exist a relativistic localization which is different from the
non-covariant Newton-Wigner localization.

\subsection{The standard case: halfinteger spin}

For simplicity we start from the Hilbert space of complex momentum space wave
function of the irreducible $(m,s=0)$ representation for a neutral
(selfconjugate) scalar particle. In this case we only need to remind the
reader of published results \cite{1997}\cite{AP}\cite{BGL}\cite{JMP}.
\begin{align}
&  H_{Wig}=\left\{  \psi(p)|\int\left|  \psi(p)\right|  ^{2}\frac{d^{3}%
p}{2\sqrt{p^{2}+m^{2}}}<\infty\right\} \\
&  \left(  \frak{u}(\Lambda,a)\psi\right)  (p)=e^{ipa}\psi(\Lambda
^{-1}p)\nonumber
\end{align}
For the construction of the real subspace $H_{R}(W_{0})$ of the standard
$t$-$z$ wedge $W_{0}=(z>\left|  t\right|  ,x,y$ arbitrary) we use the $z-t$
Lorentz boost $\Lambda_{z-t}(\chi)\equiv\Lambda_{W_{0}}(\chi)$%
\begin{equation}
\Lambda_{W_{0}}(\chi):\left(
\begin{array}
[c]{c}%
t\\
z
\end{array}
\right)  \rightarrow\left(
\begin{array}
[c]{cc}%
\cosh\chi & -\sinh\chi\\
-\sinh\chi & \cosh\chi
\end{array}
\right)  \left(
\begin{array}
[c]{c}%
t\\
z
\end{array}
\right)
\end{equation}
which acts on $H_{Wig}$ as a unitary group of operators $\frak{u}(\chi)\equiv$
$\frak{u}(\Lambda_{z-t}(\chi),0)$ and the $z$-$t$ reflection $r:$
($z,t)\rightarrow(-z$,$-t)$ which, since it involves time reflection, is
implemented on Wigner wave functions by an unti-unitary operator $\frak{u}(r)$
\cite{JMP}\cite{BGL}. One then forms (by the standard functional calculus) the
unbounded\footnote{The unboundedness is of crucial importance since the domain
of definition is the only distinguishing property of the involution (\ref{s})
into which geometric properties (causally closed regions in Minkowski space)
are encoded.} ``analytic continuation'' in the rapidity $\frak{u}%
(\chi\rightarrow i\chi)$ which leads to unbounded positive operators. Using a
notation which harmonizes with that of the modular theory (see appendix A), we
define the following operators in $H_{Wig}$
\begin{align}
&  \frak{s}=\frak{\ \frak{j}}\delta^{\frac{1}{2}}\label{pol}\\
\frak{\ \frak{j}}  &  =\frak{u}(r)\nonumber\\
&  \delta^{it}=\frak{u}(\chi=-2\pi t)\nonumber\\
&  \left(  \frak{s}\psi\right)  (p)=\psi(-p)^{\ast} \label{s}%
\end{align}
Note that all the operators are functional-analytically extended geometrically
defined objects within the Wigner theory; in particular the last line is the
action of an unbounded involutive $\frak{s}$ on Wigner wave functions which
involves complex conjugation as well as an ``analytic continuation'' into the
negative mass shell. Note that $\frak{u}(r)$ is apart from a $\pi$-rotation
around the x-axis the one-particle version of the TCP operator. The last
formula for $\frak{s}$ would look the same even if we would have started from
another wedge $W\neq W_{0}.$ This is quite deceiving since physicists are not
accustomed to consider the domain of definition as an integral part of the
definition of the operator. If the formula would describe a bounded operator
the formula would define the operator uniquely but in the case at hand
$dom\frak{s\equiv}dom\frak{s}_{W_{0}}\neq dom\frak{s}_{W}$ for $W_{0}\neq W$
since the domains of $\delta_{W_{0}}$ and $\delta_{W}$ are quite different; in
fact the geometric positions of the different $W^{\prime}s$ can be recovered
from the $\frak{s}^{\prime}s$. All Tomita S-operators are only different in
their domains but not in their formal appearance; this makes modular theory a
very treacherous subject.

The content of (\ref{pol}) is nothing but an adaptation of the spatial version
of the Bisognano-Wichmann theorem to the Wigner one-particle theory
\cite{JMP}\cite{BGL}. The former is in turn a special case of Rieffel's and
van Daele's spatial generalization \cite{Rieffel} of the operator-algebraic
Tomita-Takesaki modular theory (see appendix A). Since the antiunitary $t$-$z$
reflection commutes with the $t$-$z$ boost $\delta^{it}$, it inverts the
unbounded ($\delta^{i})^{-i}=\delta$ i.e. $\frak{j}\delta=\delta^{-1}%
\frak{j.}$ As a result of this commutation relation, the unbounded antilinear
operator $\frak{s}$ is involutive on its domain of definition i.e.
$\frak{s}^{2}\subset1$ so that it may be used to define a real subspace
(closed in the real sense i.e. its complexification is not closed) as
explained in the appendix. The definition of $H_{R}(W_{0})$ is in terms of +1
eigenvectors of $\frak{s}$
\begin{align}
H_{R}(W_{0})  &  =clos\left\{  \psi\in H_{Wig}|\,\frak{s}\psi=\psi\right\} \\
&  =clos\left\{  \psi+\frak{s}\psi|\,\psi\in dom\frak{s}\right\} \nonumber\\
\frak{si\psi}  &  =\frak{-i\psi},\text{ }\psi\in H_{R}(W_{0})\nonumber
\end{align}
The +1 eigenvalue condition is equivalent to analyticity of $\delta^{it}\psi$
in $i\pi<Imt<0$ (and continuity on the boundary) together with a reality
property relating the two boundary values on this strip. The localization in
the opposite wedge i.e. the $H_{R}(W^{opp})$ subspace turns out to correspond
to the symplectic (or real orthogonal) complement of $H_{R}(W)$ in $H_{Wig}$
i.e.%
\begin{equation}
\operatorname{Im}(\psi,H_{R}(W_{0}))=0\Leftrightarrow\psi\in H_{R}(W_{0}%
^{opp})\equiv\frak{j}H_{R}(W_{0})
\end{equation}
One furthermore finds the following properties for the subspaces called
``standardness''
\begin{align}
&  H_{R}(W_{0})+iH_{R}(W_{0})\,\,is\,\,dense\,\,in\,\,H_{Wig}\\
&  H_{R}(W_{0})\cap iH_{R}(W_{0})=\left\{  0\right\} \nonumber
\end{align}
For completeness we sketch the proof. The closedness of the densely defined
$\frak{s}$ leads to the following decomposition of the domain $dom\frak{s}$%
\begin{align}
dom(\frak{s})  &  =\left\{  \psi\in H_{Wig}|\,\psi=\frac{1}{2}\left(
\psi+\frak{s\psi}\right)  +\frac{i}{2}\left(  \psi-\frak{s\psi}\right)
\right\} \\
&  =H_{R}(W_{0})+iH_{R}(W_{0})\nonumber
\end{align}
On the other hand from $\psi\in H_{R}(W_{0})\cap iH_{R}(W_{0})$ one obtains
\begin{align}
\psi &  =\frak{s}\psi\\
i\psi &  =\frak{si\psi=-is\psi=-i\psi}\nonumber
\end{align}
from which $\psi=0$ follows. In the appendix it was shown that vice versa the
standardness of a real subspace $H_{R}$ leads to the modular objects
$\frak{j},\delta$ and $\frak{s}$.

Since the Poincar\'{e} group acts transitively on the $W^{\prime}s$ and
carries the $W_{0}$-affiliated $\frak{u}(\Lambda_{W_{0}}(\chi)),\frak{u}%
(r_{W_{0}})$ into the corresponding $W$-affiliated L-boosts and reflections,
the subspaces $H_{R}(W)\,$have the following covariance properties
\begin{align}
\frak{u}(\Lambda,a)H_{R}(W_{0})  &  =H_{R}(W=\Lambda W_{0}+a)\\
\frak{s}_{W}  &  =\frak{u}(\Lambda,a)s_{W_{0}}\frak{u}(\Lambda,a)^{-1}%
\nonumber
\end{align}

Having arrived at the wedge localization spaces, one may construct
localization spaces for smaller spacetime regions by forming intersections
over all wedges containing this region $\mathcal{O}$
\begin{equation}
H_{R}(\mathcal{O})=\bigcap_{W\supset\mathcal{O}}H_{R}(W) \label{int}%
\end{equation}
These spaces are again standard and covariant. They have their own
``pre-modular'' (see the appendix on the spatial theory, the true Tomita
modular operators appear in the next section) object $\frak{s}_{\mathcal{O}}$
$\frak{\ }$and the radial and angular part $\delta_{\mathcal{O}}$ and
$\frak{j}_{\mathcal{O}}$ in their polar decomposition (\ref{pol}), but this
time their action cannot be described in terms of spacetime diffeomorphisms
since for massive particles the action is not implemented by a geometric
transformation in Minkowski space. To be more precise, the action of
$\delta_{\mathcal{O}}^{it}$ is only local in the sense that $H_{R}%
(\mathcal{O})$ and its symplectic complement $H_{R}(\mathcal{O})^{\prime
}=H_{R}(\mathcal{O}^{\prime})$ are transformed onto themselves (whereas
$\frak{j}$ interchanges the original subspace with its symplectic complement),
but for massive Wigner particles there is no geometric modular transformation
(in the massless case there is a modular diffeomorphism of the compactified
Minkowski space). Nevertheless the modular transformations $\delta
_{\mathcal{O}}^{it}$ for $\mathcal{O}$ running through all double cones and
wedges (which are double cones ``at infinity'') generate the action of an
infinite dimensional Lie group. Except for the finite parametric Poincar\'{e}
group (or conformal group in the case of zero mass particles) the action is
partially ``fuzzy'' i.e. not implementable by a diffeomorphism on Minkowski
spacetime but still being the product of modular group action where each
factor respects the causal closure (causal ``horizon'') of a region
$\mathcal{O}$ (more precisely: it is asymptotically gemometric near the
horizon). The emergence of these \textit{fuzzy acting Lie groups is a pure
quantum phenomenon}; there is no analog in classical physics. They describe
hidden symmetries \cite{SW1}\cite{S-W3} which the Lagrangian formalism does
not expose.

Note also that the modular formalism characterizes the localization of
subspaces. In fact for the present $(m,s=0)$ Wigner representations the spaces
$H_{R}(\mathcal{O})$ have a simple description in terms of Fourier transforms
of spacetime-localized test functions. In the selfconjugate case one finds
\begin{equation}
H_{R}(\mathcal{O})=rclos\left\{  \psi=E_{m}\tilde{f}\,|f\in\mathcal{D}%
(\mathcal{O}),f=f^{\ast}\right\}  \label{supp}%
\end{equation}
where the closure is taken within the real subspace i.e. one imposes the
reality condition $f=f^{\ast}$ in the mass-shell restriction corresponding to
a projector $E_{m}$ acting on the Fourier transform i.e. ($E_{m}\tilde
{f})(p)=\left(  E_{m}\tilde{f}\right)  ^{\ast}(-p),\,p^{2}=m^{2},p_{0}>0.$
This space may also be characterized in terms of a closure of a space of
entire functions with a Pailey-Wiener asymptotic behaviour. From these
representations (\ref{int}\ref{supp}) it is fairly easy to conclude that the
inclusion-preserving maps $\mathcal{O}\rightarrow H_{R}(\mathcal{O})$ are maps
between orthocomplemented lattices of causally closed regions (with the
complement being the causal disjoint) and modulare localized real subspaces
(with the simplectic or real orthogonal complement). In particular one finds
$H_{R}(\mathcal{O}_{1}\cap\mathcal{O}_{2})=H_{R}(\mathcal{O}_{1})\cap
H_{R}(\mathcal{O}_{2}).$ The complement of this relation is called the
additivity property which is an indispensible requirement if the Global is
obtained by piecing together the Local.

The dense subspace $H(W)=H_{R}(W)+iH_{R}(W)$ of $H_{Wig}$ changes its position
within $H_{Wig}$ together with $W.$ If one would close it in the topology of
$H_{Wig}$ one would loose all this subtle geometric information encoded in the
$\frak{s}$-domains. One must change the topology in such a way that the dense
subspace $H(W)$ becomes a Hilbert space in its own right. This is achieved in
terms of the graph norm of $\frak{s}_{W}$ (for the characterization of the
$H_{R}(\mathcal{O})$ in terms of test function (\ref{supp}) one did not need
the $\frak{s}$-operator
\begin{equation}
\left(  \psi,\psi\right)  _{G\frak{s}}\equiv\left(  \psi,\psi\right)  +\left(
\frak{s}\psi,\frak{s}\psi\right)  <\infty
\end{equation}
This topology is simply an algebraic way of characterizing a Hilbert space
which consists of localized vectors only. It is easy to write down a modified
measure in which the $\frak{s}$ becomes a bounded operator
\begin{align}
\left(  \psi,\psi\right)  _{ther}  &  =\int\psi^{\ast}(\theta,p_{\perp}%
)\frac{1}{\delta-1}\psi(\theta,p_{\perp})d\theta\\
\psi(\theta,p_{\perp})  &  =\psi(p),\,\,p=(m_{eff}\cosh\theta,p_{\perp
},m_{eff}\sinh\theta)\nonumber
\end{align}
Clearly $\delta=\frak{s}^{\ast}\frak{s}$ and $1+\delta$ are bounded in this
norm. Defining the Fourier transform
\begin{equation}
f(\theta)=\frac{1}{\sqrt{2\pi}}\int\tilde{f}(\kappa)e^{i\kappa\theta}%
\end{equation}
The modification takes on the appearance of a \textit{thermal} Bose factor at
temperature $T=2\pi$ with the role of the Hamiltonian being played by the
Lorentz boost generator $K$ in $\delta=e^{-2\pi K}$ (which is the reason for
using the subscript $ther).$ In fact the Wigner one-particle theory preempts
the fact that the associated free field theory in the vacuum state restricted
to the wedge becomes thermal i.e. satisfies the KMS condition and the thermal
inner product becomes related to the two-point-function of that wedge
restricted QFT. We have taken a wedge because then the modular Hamiltonian K
has a geometric interpretation in terms of the L-boost, but the modular
Hamiltonian always exists; if not in a geometric sense then as a fuzzy
transformation which fixes the localization region and its causal complement.
Hence for any causally closed spacetime region $\mathcal{O}$ and its
nontrivial causal complement $\mathcal{O}^{\prime}$ there exists such a
thermally closed Hilbert space of localized vectors and for the wedge $W$ this
preempts the Unruh-Hawking effect associated with the geometric Lorentz boost
playing the role of a Hamiltonian (in case of $(m=0,s=$halfinteger$)$
representations this also holds for double cones since they are conformally
equivalent to wedges).

After having obtained some understanding of modular localization it is helpful
to highlight the difference between N-W and modular localization by a concrete
illustration. Consider the energy momentum density in a one-particle wave
function of the form $\psi_{f}=E_{m}f\in H_{R}(\mathcal{O})\,$\ where
$suppf\subset\mathcal{O}$, $f$ real%
\begin{align}
t_{\mu\nu}(x,\psi)  &  =\partial_{\mu}\psi_{f}(x)\partial_{\nu}\psi
_{f}(x)+\frac{1}{2}g_{\mu\nu}(m^{2}\psi_{f}(x)^{2}-\partial^{\nu}\psi
_{f}(x)\partial_{\nu}\psi_{f}(x))\label{exp}\\
&  =\left\langle f,c\left|  :T_{\mu\nu}(x):\right|  f,c\right\rangle
,\,\,\left|  f,c\right\rangle \equiv W(f)\left|  0\right\rangle \,\,\nonumber
\end{align}
where on the right hand side we used the standard field theoretic expression
for the expectation value of the energy-momentum density in a coherent state
obtained by applying the Weyl operator corresponding to the test function $f$
to the vacuum. Since $\psi_{f}(x)=\int\Delta(x-y,m)f(y)d^{4}y$ we see that the
one-particle expectation (\ref{exp}) complies with Einstein causality (no
superluminal propagation outside the causal influence region of $\mathcal{O}%
),$ but there is no way to affiliate a projector with the subspace
$H_{R}(\mathcal{O})$ or with coherent states (the real projectors appearing in
the appendix are really unbounded operators in the complex sense). We also
notice that as a result of the analytic properties of the wave function in
momentum space the expectation value has crossing properties, i.e. it can be
analytically continued to a matrix element of T between the vacuum and a
modular localized two-particle two-particle state. This follows either by
explicit computation or by using the KMS property on the field theoretic
interpretation of the expectation value. A more detailed investigation shows
that the appearance of this crossing (vacuum polarization) structure and the
absence of localizing projectors are inexorably related. This property of the
positive energy Wigner representations preempts a generic property of local
quantum physics: \textit{relativistic localization cannot be described in
terms of (complex) subspaces and projectors, rather this must be done in terms
of expectation values of local observables in modular localized states which
belong to real subspaces.}

The use of the inappropriate localization concept is the prime reason why
there have been many misleading papers on ``superluminal propagation'' in
which Fermi's result that the classical relativistic propagation inside the
forward light cone continues to hold in relativistic QFT was called into
question (for a detailed critical account see \cite{Bu-Yng}).

On the more formal mathematical level this absence of localizing projectors is
connected to the absence of pure states and minimal projectors in the local
operator algebras. The standard framework of QM and the concepts of ``quantum
computation'' simply do not apply to the local operator algebras since the
latter are of von Neumann type $III_{1}$ hyperfinite operator algebras\ and
not of the quantum mechanical type $I$. Therefore it is a bit misleading to
say that local quantum physics is just QM with the nonrelativistic Galilei
group replaced by Poincar\'{e} symmetry; these two requirements would lead to
the relativistic QM mentioned in the previous section whereas QFT is
characterized by micro-causality of observables and modular localization of
states. To avoid any misunderstanding, projectors in compact causally closed
local regions $\mathcal{O}$ of course exist, but they necessarily describe
fuzzy (non sharp) localization within $\mathcal{O}$ \cite{Hor} and the vacuum
is necessarily a highly entangled temperture state if restriceted via this
projector (in QM spatial restrictions only create isotopic representations
i.e. enhanced multiplicities but do not cause genuine entanglement or thermal behavior).

It is interesting that the two different localization concepts have aroused
passionate discussions in philosophical circles as evidenced e.g. from
bellicose sounding title as ``Reeh-Schlieder defeats Newton-Wigner'' in
\cite{Halvor}. As it should be clear from our presentation particle physics
finds both very useful, the first for causal (non-superluminal) propagation
and the second for scattering theory where only asymptotic covariance and
causality is required.

After having made pedagogical use of the simplicity of the scalar neutral case
in order to preempt some consequences of the modular aspects of QFT on the
level of the Wigner one-particle theory, it is now easy to add the
modifications which one has to make for charged scalar particles and those
with nonzero spin. The Wigner representation of the connected part of
a\ Poincar\'{e} group describes only one particle, so in order to incorporate
the antipartice which has identical Poincar\'{e} properties one just doubles
the Wigner space and defines the $\frak{j}$ and the $\frak{s}$ as follows
(still spin-less)
\begin{align}
&  \left(  \frak{j}\psi\right)  (p)=\psi^{c}(rp),\,\,\left(  \frak{s}%
\psi\right)  (p)=\psi^{c}(-p)\\
\psi(p)  &  =\left(
\begin{array}
[c]{c}%
\psi_{1}(p)\\
\psi_{2}(p)
\end{array}
\right)  ,\,\psi^{c}(p)=\left(
\begin{array}
[c]{c}%
\psi_{2}(p)^{\ast}\\
\psi_{1}(p)^{\ast}%
\end{array}
\right) \nonumber\\
\psi^{c}(p)  &  =C\psi(p)^{\ast},\,\,C=\left(
\begin{array}
[c]{cc}%
0 & 1\\
1 & 0
\end{array}
\right) \nonumber
\end{align}
It is then easy to see that $\frak{s}$ has a polar decomposition as before in
terms of $j$ and a Lorentz boost $\frak{s}=\frak{j\delta.}$ The real subspaces
resulting from closed +1 eigenstates of $\frak{s}$ are
\begin{equation}
H_{R}(W)=rclos\left\{  \psi(p)+\psi^{c}(-p)|\,\psi\in dom\frak{s}\right\}
\end{equation}
where the real closure is taken with respect to real linear combinations.
Again the subspaces $H_{R}(\mathcal{O})$ defined by intersection as in
(\ref{int}) admit a representation in terms of real closures of (mass shell
projected, two-component, C-conjugation-invariant) $\mathcal{O}$-supported
test function spaces as in (\ref{supp}).

However it would be misleading to conclude from this spinless example that
modular localization in positive energy Wigner representations theory is
always quite that simple. For nontrivial halfinteger spin massive particles
the 2s+1 component wave function transform according to
\begin{align}
\left(  \frak{u}(\tilde{\Lambda},a)\psi\right)  (p)  &  =e^{iap}D^{(s)}%
(\tilde{R}(\Lambda,p))\psi(\Lambda^{-1}p)\label{Wigner}\\
\tilde{R}(\Lambda,p)  &  =\alpha(L(p))\alpha(\Lambda)\alpha(L^{-1}%
(\Lambda^{-1}p))\nonumber\\
\alpha(L(p))  &  =\sqrt{\frac{p^{\mu}\sigma_{\mu}}{m}}\nonumber
\end{align}
Here $\alpha$ denotes the SL(2.C) covering (transformation of undotted
fundamental spinors) and $\tilde{R}(\Lambda,p)\,$\ is an element of the
(covering of the) ``little group'' which is the fixed point
subgroup\footnote{We will use the letter $R$ even in the massless case when
the little group becomes the noncompact Euclidean group.} of the chosen
reference vector $p_{R}=(m,0,0,0)$ on the $(m>0,s)$ orbit. $L(p)$ is the
chosen family of boosts which transform $p_{R}$ into a generic $p$ on the
orbit. The fixed point group for the case at hand is the quantum mechanical
rotation group i.e. $\tilde{R}(\Lambda,p)\in SU(2)$ and the $D$-operators are
representation matrices $D^{(s)}$ of $SU(2)$ obtaines by symmetrizing the
2s-fold SU(2) tensor products.

For $s=\frac{n}{2},n\,$odd$,$ the Wigner matrices $\tilde{R}(\Lambda_{W_{0}%
}(-2\pi t),p)$ enter the definition of the operator $\frak{s}$ and they
generally produce a square-root cut in the analytic strip region. As a
representative case of halfinteger spin we consider the case of a selfdual
massive $s=\frac{1}{2}$ particle. The fact that the $SU(2)$ Wigner rotation is
only pseudo-real i.e. that the conjugate representation (although being
$i\sigma_{2}$-equivalent to the defining one, there is no equivalence
transformation which makes them identical) forces us to double order deal with
selfconjugate Wigner transformation matrices%
\begin{align}
\psi_{d}  &  :=\frac{1}{2}\left(
\begin{array}
[c]{cc}%
1 & 1\\
-i & i
\end{array}
\right)  \left(
\begin{array}
[c]{c}%
\psi_{1}\\
i\sigma_{2}\psi_{2}%
\end{array}
\right)  ,\,\\
\psi_{d}  &  \rightarrow D_{d}\psi_{d},\,\,D_{d}=\left(
\begin{array}
[c]{cc}%
\operatorname{Re}D & \operatorname{Im}D\\
-\operatorname{Im}D & \operatorname{Re}D
\end{array}
\right) \nonumber
\end{align}
where $D$ denote the original $SU()$-valued Wigner transformation matrices.
Therefore the representation space will be represented by $4\times2$ component
spinor%
\begin{equation}
\Psi(p)=\left(
\begin{array}
[c]{c}%
\psi_{d}^{(1)}(p)\\
\psi_{d}^{(2)}(p)
\end{array}
\right)  \overset{C}{\longrightarrow}\Psi^{C}(p)=\left(
\begin{array}
[c]{c}%
\psi_{d}^{(2)}(p)\\
\psi_{d}^{(1)}(p)
\end{array}
\right)
\end{equation}
so that the definition for the spatial Tomita operator%
\begin{align}
\frak{s}\Psi(p)  &  =\Psi^{C}(-p)\\
H_{R}(W)  &  =\left\{  \Psi(p)|\,\frak{s}\Psi(p)=\Psi(p)\right\}
\curvearrowright\psi_{d}^{(1)}(p)=\psi_{d}^{(2)}(-p)^{\ast}\nonumber
\end{align}
complies with the conjugacy properties of the Wigner transformations. For
selfconjugate (Majorana) particles one has in addition $\psi_{1}=\psi_{2}.$

The original Wigner transformation $D$ (\ref{Wigner}) contains the t-dependent
2$\times2$ matrix which in Pauli matrix notation reads
\begin{equation}
\frac{1}{\sqrt{m}}\left(  \cosh2\pi t\cdot p^{0}\mathbf{1}-\sinh2\pi t\cdot
p^{1}\sigma_{1}+p^{2}\sigma_{2}+p^{3}\sigma_{3}\right)  ^{\frac{1}{2}}%
\end{equation}
which in the analytic continuation $t\rightarrow z$ develops a square root cut
in the would-be analytic strip $-i\pi<z<0.$ This square root cut in $D_{d}$
complicates the description of the domain $dom\frak{s.}$

The only way to retain strip analyticity in the presence of the Wigner
transformation law is to have a compensating singularity in the transformed
wave function $\Psi(\Lambda_{W_{0}}(-2\pi t)p)$ as t is continued into the
strip. This is achieved by factorizing the Wigner wave function in terms of
intertwiners $\alpha$. Let us make the following ansatz for the original
2-component Wigner wave function%
\begin{align}
&  \psi(p)=\alpha(L(p))\left(  E_{m}\Phi\right)  (p)\\
&  \alpha(L(p))=\sqrt{\frac{p^{\mu}\sigma_{\mu}}{m}}\nonumber\\
&  \tilde{R}(\Lambda,p)\alpha(L(\Lambda^{-1}p))=\alpha(L(p))\alpha
(\Lambda)\nonumber
\end{align}
where in the last line we wrote the intertwining relation for the intertwining
matrix $\alpha(L(p)).$ $\Phi_{\alpha}(x)\in\mathcal{D}(W_{0}),\alpha=1,2$ is a
two-component space of test functions with support in the standard wedge
$W_{0}.$ Such test functions whose associated Fourier transformed wave
functions projected onto the mass shell $\left(  E_{m}\Phi\right)  (p)$
obviously fulfill the strip analyticity are interpreted as (undotted) spinors
i.e. they are equipped with the transformation law%

\begin{equation}
\Phi(x)\rightarrow\alpha(\Lambda)\Phi(\Lambda x),\,\alpha(\Lambda)\in SL(2,C)
\end{equation}
The covariant (undotted) spinorial transformation law\footnote{Since here we
have to distinguish between undotted and dotted spinors, we use the notation
$\alpha(\Lambda)$ and $\beta(\Lambda)=\overline{\alpha(\Lambda)}$ instead of
the previous $\tilde{\Lambda}.$} changes the support in a geometric way. As a
consequence of \ group theory, the spinor wave function defined by (with
$E_{m}$ a mass shell projector as before and $\frak{u}(p)$ intertwiner matrix
$\frak{u}(p)=$ transforms according to Wigner as
\begin{equation}
\psi(p)\rightarrow\alpha(\tilde{R}(\Lambda,p))\frak{u}(\Lambda^{-1}%
p)(E_{m}\Phi)(\Lambda^{-1}p)=\frak{u}(p)\alpha(\Lambda)\psi(\Lambda^{-1}p)
\end{equation}
where in the second line we wrote the intertwining law of $\frak{u}%
(p)=\alpha(L(p))$ of which the first line is a consequence. We see that the
product Ansatz $\psi=uE_{m}\Phi$ solves the problem of the strip analyticity
since the $\frak{u}(p)$ factor develops a square root cut which compensates
that of the Wigner rotation and $E_{m}\Phi$ is analytic from the wedge
localization of $\Phi$. The test function space provides a dense set in
$H_{R}(W)$ so by adding limits, one obtains all of $H_{R}(W)$ i.e. all the
full +1 eigenspace of $\frak{s.}$ In fact this Ansatz avoids the occurance of
singular pre-factor for any causally complete localization region
$\mathcal{O};$ in the compact case the closure of the test function space
turns out to be a space of entire functions with an appropriate
Pailey-Wiener-Schwartz asymptotic behaviour reflecting the size of the double
cones $\mathcal{O}.$ Although our analyticity discussion was done on the
original Wigner representation, it immediatly carries over to the doubled
version which we have used for the construction of the real modular subspaces
$H_{R}(W).$ Again $H(W)=H_{R}(W)+iH_{R}(W)$ will be dense in $H_{Wig}$ for the
same reason as in the cases before. To obtain the solution for arbitrary
halfinteger spin one only has to use symmetrized tensor representations of
$SL(2,C)$ and its $SU(2)$ subgroup.

If we now try to represent our $\frak{s}$-operator as $j\Delta^{\frac{1}{2}}%
$in terms of geometrically defined reflections and boosts we encounter a
surprise; the geometrically defined object is different by a phase factor $i.$
This factor results from the analytically continued Wigner rotation in the
boost parameter for all halfinteger spins. The only way to compensate it
consistent with the polar decomposition is to say that the $j$ deviates from
the geometric $j_{0}$ by a phase factor $t$%
\begin{equation}
j=tj_{0},\,\,t=i
\end{equation}
It turns out that this also happens for the exeptional Wigner representations;
for d=1+2 anyons one obtains a phase factor related to the spin of the anyon
whereas for the d%
$>$%
1+3 spin towers $t$ is an operator in the infinite tower space related to the
analytically continued infinite dimensional Wigner matrix. These cases are
characteized by the failure of compact modular localization (see below).

The modular localization in the massless case is similar as long as the
helicity stays finite (trivially represented Euclidean ``translations'') is
similar. The concrete determination of the $\Lambda,p$-dependent $\tilde{R}$
requires a selection of a family of boosts i.e. of Lorentz transformations
$\tilde{L}(p)$ which relate the reference vector $p_{R}$ uniquely a general
$p$ on the respective orbit. The natural choice for the associated $2\times2$
matrices in case of d=1+3 is (we use $\alpha$ for the $SL(2,C)$
representation)
\begin{equation}
\alpha(\tilde{L}(p)_{0})=\frac{1}{\sqrt{p_{0}+p_{3}}}\left(
\begin{array}
[c]{cc}%
p_{0}+p_{3} & p_{1}-ip_{2}\\
0 & 1
\end{array}
\right)  ,\,m=0 \label{prev}%
\end{equation}
with the associated little groups being $SU(2)$ or for m=0 $\tilde{E}%
(2)\,$\ (the 2-fold covering of the 2-dim. Euclidean group)
\begin{equation}
\tilde{E}(2):\left(
\begin{array}
[c]{cc}%
e^{i\frac{\varphi}{2}} & z=a+ib\\
0 & e^{-i\frac{\varphi}{2}}%
\end{array}
\right)  ,\,\,\,m=0
\end{equation}
For the standard (halfinteger helicity) massless representations the
``z-translations'' are mapped into the identity. As a result of the projection
property of the reference vector there exists a projected form of the
intertwining relation ($\alpha(\tilde{L}(p))$ as in (\ref{prev}))
\begin{align}
&  \textsl{p}_{R}\tilde{R}(\Lambda,p)=\textsl{p}_{R}\tilde{R}(\Lambda
,p)_{11}\\
&  \tilde{R}(\Lambda,p)=\alpha(\tilde{L}(p))\tilde{\Lambda}\alpha(\tilde
{L}^{-1}(\Lambda^{-1}p))\nonumber
\end{align}
This projection allows to incorporate the one-component formalism into the
SL(2,C) matrix formalism. In fact this embedding permits to use the same mass
independent $W$-supported test function spaces as before, one only has to
replace the $E_{m}$ projectors by projectors on the zero mass orbit. Again the
definition of $\frak{j}$ generally demands a further doubling of the test
function. At the end one obtains a representation of modular localization
spaces $H_{R}(W)$ (and more generally $H_{R}(\mathcal{O})$ for double cones
$\mathcal{O}$) in terms of $W$ or $\mathcal{O}$ supported spinorial test
function spaces whose nontriviality is secured by the classical Schwartz
distribution theory.

It is easy to see that the modular formalism also works for halfinteger spin
in d=1+2 dimensions. In this case one can work with the same $2\times2$ matrix
model, we only have to restrict $SL(2,R)$ to $SL(2,R)\simeq SU(1,1)$ which is
conveniently done by omitting the $\sigma_{2}$ Pauli matrix. Choosing again
the rest frame reference vector we obtain
\begin{align}
\tilde{L}(p)  &  =+\sqrt{\frac{p^{\mu}\sigma_{\mu}}{m}},\,m>0,\text{ }%
\sigma_{2}\text{ }omitted\\
\tilde{L}(p)  &  =\frac{1}{\sqrt{p_{0}+p_{3}}}\left(
\begin{array}
[c]{cc}%
p_{0}+p_{3} & p_{1}\\
0 & 1
\end{array}
\right)  ,\,m=0\nonumber
\end{align}
with the little group $G_{l}$ being the abelian rotation or the abelian
``translation'' group respectively.
\begin{align}
\textsl{gp}_{R}\textsl{g}^{\ast}  &  =\textsl{p}_{R}\\
G_{l}  &  :\textsl{g}=\left(
\begin{array}
[c]{cc}%
\cos\frac{1}{2}\Omega & \sin\frac{1}{2}\Omega\\
\sin\frac{1}{2}\Omega & \cos\frac{1}{2}\Omega
\end{array}
\right)  ,\,m>0\nonumber\\
G_{l}  &  :\textsl{g}=\left(
\begin{array}
[c]{cc}%
1 & b\\
0 & 1
\end{array}
\right)  ,\,m=0\nonumber
\end{align}
In order to preserve the analogy in the representations, we take halfinteger
spin representations in the first case and trivial representation of the
little group in the massless case. Whereas the massless case has a modular
wedge structure like the scalar case, the modular structure of the (m,s) case
is solved by a u-intertwiner as in the previous d=1+3 case. We have and will
continue to refer to these representations with finite (half)integer finite
spin as ``standard''. Their modular localization spaces $H_{R}(\mathcal{O})$
can be described in terms of classical $\mathcal{O}$-supported test functions.
The remaining cases, here called ``exceptional'', will be treated in the next
subsection. They include the d=1+2 ``anyonic'' spin of massive particles as
well as massless cases with faithful representations of the little group in
any spacetime dimension $d\geq1+2.$ For $d\geq1+3$ they are identical to the
famous Wigner spin towers where infinitely many spins (like in a dynamical
string) are combined in one irreducible representation.

We will see that for these exceptional representations the best possible
modular localization is noncompact and generally not susceptible to a
classical description in terms of support properties of functions. This
preempts the more noncommutative properties of the associated QFTs which are
outside of Lagrangian quantization.

\subsection{Exceptional cases: anyons and infinite ``spin towers''}

The special role of d=1+2 spacetime dimensions for the existence of braid
group statistics is due to the fact that the universal covering is infinite
sheeted and not two-fold as considered in the previous section. The fastest
way to obtain a parametrization of the latter is to use the Bargmann
\cite{Barg} parametrization%
\begin{equation}
\left\{  \left(  \gamma,\omega\right)  |\,\gamma\in\mathbb{C},\left|
\gamma\right|  <1,\,\omega\in\mathbb{R}\right\}
\end{equation}
for the two-fold matrix covering
\begin{equation}
\frac{1}{\sqrt{1-\gamma\bar{\gamma}}}\left(
\begin{array}
[c]{cc}%
e^{i\frac{\omega}{2}} & \gamma e^{i\frac{\omega}{2}}\\
\bar{\gamma}e^{-i\frac{\omega}{2}} & e^{-i\frac{\omega}{2}}%
\end{array}
\right)
\end{equation}
It is then easy to abstract the multiplication law for the universal covering
from this matrix model
\begin{align}
&  \left(  \gamma_{2},\omega_{2}\right)  \left(  \gamma_{1},\omega_{1}\right)
=(\gamma_{3},\omega_{3})\\
&  \gamma_{3}=\frac{\left(  \gamma_{1}+\gamma_{2}e^{-i\frac{\omega}{2}_{1}%
}\right)  }{\left(  1+\gamma_{2}\bar{\gamma}_{1}e^{-\frac{\omega}{2}_{1}%
}\right)  }\nonumber\\
e^{i\frac{\omega_{3}}{2}}  &  =e^{i\frac{\omega_{1}+\omega_{2}}{2}}\left(
\frac{1+\gamma_{2}\bar{\gamma}_{1}e^{-i\frac{\omega}{2}_{1}}}{1+\gamma
_{2}\gamma_{1}e^{i\frac{\omega}{2}_{1}}}\right)  ^{\frac{1}{2}}\nonumber
\end{align}
>From these composition laws one may obtain the irreducible transformation law
of a (m,s)Wigner wave functions in terms of a one-component representation
involving a Wigner phase $\varphi((\gamma,\omega),p)$

But there are some quite interesting and physically potentially important
positive energy representations for which the above covariantization does not
work and the $H_{R}(\mathcal{O})$ do not have such a geometric description
i.e. the modular localization is more ''quantum'' than geometric. These
exceptional representations include $d=1+2$ spin$\neq$halfinteger anyons and
the still somewhat mysterious $d\geq1+3$ massless ``infinite spin-tower''
(called ``continuous spin'' by Wigner, unfortunately a somewhat misleading
name). These are the cases which also resist Lagrangian quantization attempts.
However the modular localization method reveal for the first time that those
representations do not allow a compact (with pointlike as limiting case)
localization in fact these cases are only consistent with a noncompact modular
localization which extends to infinity. The associated multiparticle spaces do
not have the structure of a Fock space and the localized operators describing
creation and annihilation are too noncommutative for a Lagrangian quantization interpretation.

Before we look at those special cases let us note that the localization in
wedges and in certain special intersection of two wedges is a general property
of all positive energy representations of $\mathcal{P}_{+}$. The above proof
of standardness of the $\frak{s}$ operator only uses general properties of the
boost and the r reflection which are evidently true in each positive energy
representation of the extended Poincar\'{e} group $\widetilde{\mathcal{P}}%
_{+}.$ A bit more tricky is the nontriviality of the following intersected spaces

\begin{theorem}
(Guido and Longo \cite{GL}) Let W$_{1}$ and W be orthogonal wedges (in the
sense of orthogonality of their spacelike edges) and define $W_{2}=\Lambda
_{W}(-2\pi t)W_{1}.$ Then $H(W_{1}\cap W_{2})\equiv H_{R}(W_{1}\cap
W_{2})+iH_{R}(W_{1}\cap W_{2})$ is dense in the positive energy representation
space $\mathcal{P}_{+}.$
\end{theorem}

The size of the intersection decreases with increasing t. It is conic with
apex at the origin, but it does not look like a spacelike cone since it
contains lightlike rays (for t$\rightarrow\infty$ its core is a lightlike string).

\begin{proof}
>From the assumptions one obtains a geometric expression for $\frak{s}%
_{2}\frak{s}_{1}$%
\[
\frak{s}_{2}\frak{s}_{1}=\Delta_{W_{2}}^{it}\Delta_{W_{1}}^{-\frac{1}{2}%
}\Delta_{W_{2}}^{it}\Delta_{W_{1}}^{\frac{1}{2}}%
\]
where we used the orthogonality assumption via $\frak{j}_{W_{1}}\Delta_{W_{2}%
}^{it}$ $\frak{j}_{W_{1}}=\Delta_{W_{2}}^{-it}.$ The claimed density is
equivalent to the denseness of the subspace:
\[
\left\{  \psi|\,\frak{s}_{2}\frak{s}_{1}\psi=\psi\right\}  \Leftrightarrow
\left\{  \psi|\,\Delta_{W_{1}}^{-\frac{1}{2}}\Delta_{W_{2}}^{it}\Delta_{W_{1}%
}^{\frac{1}{2}}\psi=\Delta_{W_{2}}^{-it}\psi\right\}
\]
but according to a theorem in \cite{GL} this is a consequence of the denseness
of the domain of $\Delta_{W_{1}}^{-\frac{1}{2}}\Delta_{W_{2}}^{it}%
\Delta_{W_{1}}^{\frac{1}{2}}$ which holds for every unitary representation of
SL(2,R) which, as easily shown, is the group generated by the two orthogonal wedges.
\end{proof}

Before this theorem will be applied to the localization of the exceptional
Wigner representation it is instructive to recall the argument for the lack of
compact localization in these cases.

Any localization beyond those of group theoretical origin requires the
construction of at least partial intertwiners. Before we comment on this let
us first show that in the cases of d=1+2 anyonic and d=1+3 infinite spin a
compact localization is impossible (which also shows that there are no
intertwiners in the previous sense). The typical causally closed simply
connected compact region has the form of a double cone i.e. the intersection
of the upper light cone with the lower one. Since in terms of wedges one needs
infinitely many intersections, we will prove the even the larger region of the
intersection of two wedges (which is infinite in transverse direction) has a
trivial $H_{R}.$

In order to compute the action of $\frak{s}$ we use the Wigner cocycle
(\ref{Wigner}) for the t-x boost $\Lambda_{W_{0}}$
\begin{align}
e^{is\Omega(\Lambda_{W_{0},},p)}  &  =\left(  \frac{1-\gamma(p)\gamma
_{t}+\left(  \gamma_{t}-\gamma(p)\right)  \overline{\gamma(\Lambda_{W_{0}%
}(-t)p)}}{c.c.}\right)  ^{s}\\
&  =u(p)u(\Lambda_{W_{0}}(-t)p),\,\,u(p)\equiv(\frac{p_{0}-p_{1}+m+ip_{2}%
}{p_{0}-p_{1}+m-ip_{2}})^{s}\nonumber
\end{align}
This formula results by specialization from the following formula for the
action of the L-group on one-component massive Wigner wave functions
\cite{Mu-S}\cite{Mund}
\begin{align*}
&  \left(  \frak{u}\psi\right)  (p,s)=e^{is\Omega(\tilde{R}(\Lambda,p))}%
\psi(\Lambda^{-1}p)\\
&  e^{is\Omega(\Lambda(\omega,\gamma),p)}=e^{is\frac{\omega}{2}}\left(
\frac{1-\gamma(p)\bar{\gamma}e^{-i\frac{\omega}{2}}+(\gamma-\gamma
(p)\bar{\gamma}e^{-i\frac{\omega}{2}})\bar{\gamma}(\Lambda(\gamma,\omega
)^{-1}p))}{c.c.}\right)  ^{s}%
\end{align*}
and a similar phase factor for the massless case with a faithful little group representation.

In case of the d=1+3 massless spin-tower representation this is more tricky.
One finds
\begin{align}
\left(  \frak{u}(\Lambda,a)\psi\right)  (p)  &  =e^{iap}V_{\Xi,\pm}(\tilde
{R}(\Lambda,p))\psi(\Lambda^{-1}p)\\
\left(  V_{\Xi,\pm}(\Lambda_{z,\varphi})f\right)  (\theta)  &  =\left\{
\begin{array}
[c]{c}%
\left\{  \exp i(\Xi\left|  z\right|  \cos(\arg z-\vartheta))\right\}
f(\vartheta-\varphi)\\
\left\{  \exp i(\Xi\left|  z\right|  \cos(\arg z-\vartheta)+\frac{1}{2}%
\varphi)\right\}  f(\vartheta-\varphi)
\end{array}
\right. \nonumber
\end{align}
with the + sign corresponding to an integer valued spin tower. In this case
the infinite component wave function $\psi(p)$ is a square integrable map from
the momentum space mass shell to functions with values in the L$_{2}$ space on
the circle (in which the noncompact $\tilde{E}(2)$ group is irreducibly
represented by the last formula). $\Xi$ is an invariant (Euclidean ``mass'')
of the $\tilde{E}(2)$ $\,$representation$.$ Scaling the $\Xi$ to one and
introducing a ``spin basis'' (discrete Fourier-basis) $e^{in\varphi}$, the
$V_{\Xi,\pm}(\Lambda_{\varphi})$ becomes diagonal and the translational part
$V_{\Xi,\pm}(\Lambda_{z})$ can be written in terms of Bessel functions
\begin{equation}
V_{\Xi,\pm}(\Lambda_{z})_{n,m}=\left(  \frac{z}{\left|  z\right|  }\right)
^{n-m}J_{n-m}(\Xi\left|  z\right|  )
\end{equation}
>From this one can study the analyticity behavior needed for the modular localization.

The following theorem may is easily established

\begin{theorem}
The d=1+2 representations with s$\neq$halfinteger and the d=1+3 Wigner spin
tower representations do not allow a compact double cone localization.
\end{theorem}

For the spin tower this was already suggested by an ancient No-Go theorem of
Yngvason \cite{Yng} who showed that there is an incompatibility with the
Wightman setting. We will prove in fact the slightly stronger statement that
the space $H_{R}(W\cap W_{a}^{^{\prime}})$ which describes the intersection of
a wedge with its translated opposite (which has still a noncompact transversal
extension) is trivial. This implies a fortiori the triviality of compact
double cone intersections. The common origin of the weaker localization
properties for the exceptional positive energy representations is the fact
that the analytical continuation of the wave function to the opposite boundary
of the strip (which combines together with the action of the
charge-conjugating geometric involution to a would be $\frak{s}$) has in
addition a matrix part (a phase factor for d=1+2) which has to be cancelled by
a compensating modification of the involution part
\begin{equation}
\frak{j=tj}_{geo}%
\end{equation}
The $t$, which in the case of the spin-tower is a complicated operator in the
representation space of the little group, is the preempted field theoretic
twist operator T whose presence shows up in commutation relations of spacelike
(noncompactly) localized operators (braid group statistics in case of d=1+2).

According to the second last theorem the localization in the noncompact
intersection of two wedges in a selected relative position (where the second
one results from applying an ``orthogonal'' boost to the first) is always
possible for all positive energy representations in all spacetime dimensions.
But only in d=1+2 this amounts to a spacelike cone localization (with a
semiinfinite spacelike string as a core). In that case one knows that
plektonic situations do not allow for a better localization. However there is
a problem with the application of that theorem to anyons since it refers to
the representation of the Poincare group in $d\geq3$ spacetime but not to its
covering $\mathcal{\tilde{P}}_{+}$ in $d=3$ which would be needed for the case
of anyons. Fortunately Mund has found a direct construction of spacelike cone
$C$ localized subspaces $H_{R}(C)$ in terms of a partial intertwiner $u(p)$
and subspace of of doubled test functions $\Phi$ with supports in spacelike
cones. If one starts from the standard $x$-$t$ wedge and wants to localize in
cones which contain the negative y-axis then Mund's localization formula and
his partial $u$ (to be distinguished from the previous $\frak{u}$) are%
\begin{equation}
u(p)E_{m}\Phi,\,\,u(p)=(\frac{p^{0}-p^{1}}{m})^{s}(\frac{p_{0}-p_{1}+m+ip_{2}%
}{p_{0}-p_{1}+m-ip_{2}})^{s} \label{spread}%
\end{equation}
For spacelike cones along other axis the form of the partial intertwiner
changes. Running through all $C$-localized test functions the formula
describes a dense set of spacelike cone-localized Wigner wave function only
for those spacelike cones which contain the negative y-axis after apex($C$)
has been shifted to the origin (which includes the standard $x$-$t$ wedge as a
limiting case). He then shows an interesting ``spreading'' mechanism namely
that if one chooses a better localized function with compacr support in that
region, the effect of the partial intertwiner `` is to radially extend the
support to spacelike infinity. The anyonic spin Wigner representation can be
encoded into many infinite dimensional covariant representations \cite{Mu-S}
(also appendix), but this does not improve the localization since infinite
dimensional covariant transformation matrices, unlike finite dimensional ones,
are not entire functions of the group parameters.

For d=1+3 the intersection region has at its core a 2-dimensional spacelike
half-plane. There is good reason to believe that this is really the optimally
possible localization for the spin-tower representation. The argument is based
on converting this representation into the factorizing form $uE_{m}f$ where
$u$ is the infinite dimensional intertwiner from the covariant representation
(appendix) to the Wigner representation. The best analytic behavior which the
unitary representation theory of the L-group (necessarily infinite
dimensional) can contribute to modular localization seems to be that of the
above Guido-Longo theorem. Whereas for the standard representations the
support of the classical test function multiplets determine the best
localization region (because the finite dimensional representations of the
Lorentz group are entire analytic functions), the exceptional representations
spread any test function localization which tries to go beyond those which
pass through the intertwiner. This goes hand in hand with a worsening of the
spacelike commutativity properties in the associated operator algebras.
Therefore in the case in which the modular localization cannot be encoded into
the support property of a test function multiplet, we often use the word
``quantum localization''. \ These are the cases which cannot not be described
as a quantized classical structure or in terms of Euclidean functional integrals.

As will be shown in the next section the QFT associated with such particles do
not allow sub-wedge PFGs i.e. better than wedge-localized operators which
applied to the vacuum create one-particle states free of vacuum polarization.

Whereas in standard Boson/Fermion systems (halfinteger spin representations)
the vacuum polarization is caused by the interaction (this can be used to
define the intrinsic meaning of interaction for such systems), the sub-wedge
vacuum polarization phenomenon associated with the QFT of the exceptional
Wigner representations is of a more kinematical kind; it occurs in those other
cases already without interaction; the polarization clouds are simply there to
sustain e.g. the anyonic spin\&statistics connection.

\section{From Wigner representations to the associated local quantum physics}

In the following we will show that such net of operator algebras of free
particles with halfinteger spin/helicity can be directly constructed from the
net of modular localized subspaces in standard Wigner representations. For
integral spin $s$ one defines with the help of the Weyl functor $Weyl(\cdot)$
the local von Neumann algebras \cite{Sch1}\cite{BGL} generated from the Weyl
operators as
\begin{equation}
\mathcal{A}(W):=alg\left\{  Weyl(f)|f\in H_{R}(W)\right\}  \label{Weyl}%
\end{equation}
a process which is sometimes misleadingly called ``second quantization''.
These Weyl generators have the following formal appearance in terms of Wigner
(momentum space) creation and annihilation operators and modular localized
wave functions
\begin{align}
&  H_{R}(W)\overset{\Gamma}{\rightarrow}Weyl:\,\,f\rightarrow
Weyl(f)=e^{iA(f)}\\
&  A(f)=\sum_{s_{3}=-s}^{s}\int(a^{\ast}(p,s_{3})f_{s_{3}}(p)+b^{\ast}%
(p,s_{3})f_{s_{3}}^{\ast}(-p)+h.c.)\frac{d^{3}p}{2\omega}\nonumber
\end{align}
It is helpful to interprete the operator $A(f)$ as an inner product%
\begin{equation}
A(f)=\int\left(
\begin{array}
[c]{cc}%
a^{\ast}(p) & b^{\ast}(p)
\end{array}
\right)  \left(
\begin{array}
[c]{c}%
f(p)\\
f^{\ast}(-p)
\end{array}
\right)  \frac{d^{3}p}{2\omega}+h.c \label{Segal}%
\end{equation}
of an operator bra with a ket vector of a 2$\times(2s+1)$ eigenfunction of
$\frak{s}$ representing a vector in $H_{R}(W).$ The formula refers only to
objects in the Wigner theory; covariant fields or wave functions do not enter
here. Unlike those covariant objects, the Weyl functor is uniquely related to
the (m,s) Wigner representation. The special hermitian combination entering
the exponent of the Weyl functor is sometimes called the I. Segal operator
\cite{Segal}.

The local net $\left\{  \mathcal{A}(\mathcal{O})\right\}  _{\mathcal{O}%
\in\mathcal{K}}$ may be obtained in two ways, either one first constructs the
spaces $H_{R}(\mathcal{O})$ via (\ref{int}) and then applies the Weyl functor,
or one first constructs the net of wedge algebras (\ref{Weyl}) and then
intersects the algebras according to
\begin{equation}
\mathcal{A(O)}=\bigcap_{W\supset\mathcal{O}}A(W)
\end{equation}
The proof of the net properties follows from the well-known theorem that the
Weyl functor relates the orthocomplemented lattice of real subspaces of
$H_{Wig}$ (with the complement $H_{R}^{\prime}$ of $H_{R\text{ }}$being
defined in the symplectic sense of the imaginary part of the inner product in
$H_{Wig})$ to von Neumann subalgebras $\mathcal{A}(H_{R})\subset
\mathcal{B}(H_{Fock})$

This functorial mapping $\Gamma$ also maps the above pre-modular operators
into those of the Tomita-Takesaki modular theory
\begin{equation}
J,\Delta,S\frak{=\Gamma(\ \frak{j},\delta,s)}%
\end{equation}
Whereas the pre-modular operators of the spatial theory (denoted by small
letters) act on the Wigner space, the modular operators $J,\Delta$ have an
$Ad$ action ($AdUA\equiv UAU^{\ast}$) on von Neumann algebras in Fock space
which makes them objects of the Tomita-Takesaki modular theory
\begin{align}
&  SA\Omega=A^{\ast}\Omega,\,S=J\Delta^{\frac{1}{2}}\\
&  Ad\Delta^{it}\mathcal{A}=\mathcal{A}\nonumber\\
&  AdJ\mathcal{A}=\mathcal{A}^{\prime}\nonumber
\end{align}
The operator $S$ is that of Tomita i.e. the unbounded densely defined normal
operator which maps the dense set $\left\{  A\Omega|\,A\in\mathcal{A}%
(W)\right\}  $ via $A\Omega\rightarrow A^{\ast}\Omega$ into itself and gives
$J$ and $\Delta^{\frac{1}{2}}$ by polar decomposition. The nontrivial
miraculous properties of this decomposition are the existence of an
automorphism $\sigma_{\omega}(t)=Ad\Delta^{it}$ which propagates operators
within $\mathcal{A}$ and only depends on the state $\omega$ (and not on the
implementing vector $\Omega)$ and a that of an antiunitary involution $J$
which maps $\mathcal{A}$ onto its commutant $\mathcal{A}^{\prime}.$ The
theorem of Tomita assures that these objects exist in general if $\Omega$ is a
cyclic and separating vector with respect to $\mathcal{A}.$

An important thermal aspect of the Tomita-Takesaki modular theory is the
validity of the Kubo-Martin-Schwinger (KMS) boundary condition \cite{Haag}
\begin{equation}
\omega(\sigma_{t-i}(A)B)=\omega(B\sigma_{t}(A)),\,\,A,B\in\mathcal{A}
\label{KMS}%
\end{equation}
i.e. the existence of an analytic function $F(z)\equiv\omega(\sigma_{z}(A)B)$
holomorphic in the strip $-1<Imz<0$ and continuous on the boundary with
$F(t-i)=\omega(B\sigma_{t}(A))$ or briefly (\ref{KMS})$.$ The fact that the
modular theory applied to the wedge algebra has a geometric aspect (with $J$
equal to the TCP operator times a spatial rotation and $\Delta^{it}%
=U(\Lambda_{W}(2\pi t))$) is not limited to the interaction-free theory
\cite{Haag}. These formulas are identical to the standard thermal KMS property
of a temperature state $\omega$ in the thermodynamic limit if one formally
sets the inverse temperature $\beta=\frac{1}{kT}$ equal to $\beta=-1.\,$This
thermal aspect is related to the Unruh-Hawking effect of quantum matter
enclosed behind event/causal horizons.

For halfinteger spin, the Weyl functor has to be replaced by the Clifford
functor $R$. In the previous section we already noted that there exists a
mismatch between the geometric and the spatial complement which led to the
incorporation of an additional phase factor $i$ into the definition of
$\frak{j}.$

A Clifford algebra is associated to a real Hilbert space $H_{R}$ with generators%

\begin{align}
&  R:\mathcal{S}(\mathbb{R}^{4})\rightarrow B(H_{R})\\
&  \left(  f,g\right)  _{R}=\operatorname{Re}\left(  f,g\right) \nonumber
\end{align}
where the real inner product is written as the real part of a complex one. One sets%

\begin{equation}
R^{2}(f)=(f,f)_{R}\mathbb{I}%
\end{equation}
or%

\begin{equation}
\{R(f),R(g)\}=2(f,g)_{R}\mathbb{I}%
\end{equation}
where $\mathcal{S}(\mathbb{R}^{4})$ is the Schwartz space of test functions
over $\mathbb{R}^{4}$ and $B(H_{R})$ is the space of bounded operators over
$H_{R}$.

These $R(f)$'s generates $Cliff(H_{R})$ as polynomials of $R$'s. The norm is
uniquely fixed by the algebraic relation, e.g.%

\begin{equation}
||R(f)||^{2}=|R(f)^{\ast}R(f)||-||R^{2}(f)||=||f||_{R}%
\end{equation}
and similarly for all polynomials, i.e., on all $Cliff(H_{R})$. The norm
closure of the Clifford algebra is sometimes called $CAR(H_{R})$ (canonical
anti-commutation) C$^{\ast}$-algebra. It is unique (always up to C$^{\ast}%
$-isomorphisms) and has no ideals. This Clifford map may be used as the analog
of the Weyl functor in the case of halfinteger spin.

It turns out to be more useful to work with a alternative version of $CAR$
which is due to Araki: the selfdual $CAR$-algebra. In that description, the
reality condition is implemented via a antiunitary involution $\Gamma$ inside
the larger complex Hilbert space $H$. Now%

\begin{align}
f  &  \longrightarrow B(f)\\
B(f)^{\ast}  &  =B(\Gamma f)\nonumber\\
\{B^{\ast}(f),B(g)\}  &  =(f,g)\mathbb{I}\nonumber
\end{align}
is a complex linear map of $H$ into generators a normed *-algebra whose
closure is by definition the C*-algebra $CAR(K,\Gamma)$. The previous Clifford
functor results from the selfadjoint objects $B(\Gamma f)=B(f)$ or $\Gamma
f=f.$ In physical terms $\Gamma$ is the charge conjugation operation $C$ which
enters the definition of the $\frak{s}$-operator. The functor maps this
spatial modular object into an operator of the Clifford algebra; the analog of
(\ref{Segal}) is
\begin{equation}
f\in H_{R}(W)\rightarrow R(f)=\Psi\cdot f+h.c.
\end{equation}
where, as explained in section 2.2, the Wigner wave function $f\in H_{R}(W)$
interpreted as a $4\times(2s+1)$ component column vector and $\Psi$ is a bra
vector of Wigner creation and annihilation operators. As a consequence of the
presence of a twist factor in the spatial involution $j=tj_{geo}$ one obtains
a twist operator in the algebraic involution $J$
\begin{align}
S  &  =J\Delta^{\frac{1}{2}},\,\,J=TJ_{geo}\\
T  &  =\frac{1-iU(2\pi)}{1-i}=\left\{
\begin{array}
[c]{c}%
1\,\,on\text{\thinspace\thinspace}even\\
i\,\,on\,\,odd
\end{array}
\right. \nonumber\\
SA\Omega &  =A^{\ast}\Omega,\,\,A\in\mathcal{A}(W)=a\lg\left\{  B(f)|\,f\in
H_{R}(W)\right\} \nonumber
\end{align}
The presence of the twist operator (which is one on the even and $i$ on the
odd subspaces of $H_{Fock}$) accounts for the difference between the von
Neumann commutant $\mathcal{A}(W)^{\prime}$ and the geometric opposite
$\mathcal{A}(W^{\prime}).$ The bosonic CCR (Weyl) and the fermionic CAR
(Clifford) local operator algebras are the only ones which permit a functorial
interpretation in terms of a ``quantization'' of classical function algebras.
In the next section we will take notice of the fact that they are also the
only QFTs which possess sub-wedge-localized PFGs.

In the case of d=1+2 anyonic spin representations the presence of a plektonic
twist has the more radical consequences. Whereas the fermionic twist is still
compatible with the existence of PFGs and free fields in Fock space, the twist
associated with genuine braid group statistics causes the presence of vacuum
polarization for any sub-wedge localization region. The same consequences hold
for the spin tower representations. .

Our special case at hand, in which the algebras and the modular objects are
constructed functorially from the Wigner theory, suggest that the modular
structure for wedge algebras may always have a geometrical significance
associated with a fundamental physical interpretation in any QFT. This is
indeed true, and within the Wightman framework this was established by
Bisognano and Wichmann \cite{Haag}. In the general case of an interacting
theory in d=1+3 with compact localization (which according to the DHR theory
is necessarily a theory of interacting Bosons/Fermions) the substitute for a
missing functor between a spatial and an algebraic version of modular theory
is the modular map between a real subspace of the full Hilbert space $H$ and a
local subalgebra of algebra of all operators $B(H).$ In a theory with
asymptotic completeness i.e. with a Fock space incoming (outgoing) particle
structure $H=H_{Fock}$ the scattering operator $S_{scat}$ turns out to play
the role of a relative modular invariant between the wedge algebra of the free
incoming operators and that of the genuine interacting situation
\begin{align}
J  &  =J_{0}S_{scat}\\
S  &  =S_{0}S_{scat}%
\end{align}
This relation follows directly by rewriting the TCP transformation of the
S-matrix and the use of the relation of $J$ with the TCP operator. The
computation of the real subspaces $H_{R}(W)\in H_{Fock}$ requires
diagonalization of the S-matrix. The difficult step about which presently
nothing is known is the passing from these subspaces to wedge-subalgebras
whose selfadjoint part applied to the vacuum generate these subspaces.
Although it is encouraging that the solution of the inverse problem
$S_{scat}\rightarrow\,\left\{  \mathcal{A}(\mathcal{O})\right\}
_{\mathcal{O}\in\mathcal{K}}$ is unique \cite{S-matrix}, a general formalism
which takes care of the existence part of the problem is not known apart from
some special but very interesting cases which will be presented in the next
section. Connes has developed a theory involving detailed properties of the
natural modular cones $\mathcal{P}_{\mathcal{A}(W),\Omega}$ which are
affiliated with a single standard pair $(\mathcal{A}(W),\Omega)$ (the net
structure is not used) but it is not clear how to relate his facial conditions
on these cones to properties of local quantum physics. As a matter of fact
even in the case of standard Wigner representations it is not clear how one
could obtain the modular algebraic structure if one would be limited to the
Connes method \cite{Connes} without the functorial relation. For these reasons
the modular based approach which tries to use the twist/S-matrix factor in
$J=J_{0}T$ respectively $J=J_{0}S_{scat}$ for the determination of the
algebraic structure of $\mathcal{A}(W)$ and subsequently computes the net
$\,\left\{  \mathcal{A}(\mathcal{O})\right\}  _{\mathcal{O}\in\mathcal{K}}$ by
forming intersections is presently limited to theories which permit only
vitual but no real particle creation. Besides the exeptional Wigner
representation (anyons, spin towers) which lead to a twist and changed
spacelike commutation relations, the only standard (bosonic, fermionic)
interacting theories are the $S_{scat}=S_{el}$ models of the d=1+1
bootstrap-formfactor setting (factorizing models).

For those readers who are familiar with Weinberg's method of passing from
Wigner representation to covariant pointlike free fields, it may be helpful to
add a remark which shows the connection to the modular approach. For writing
covariant free fields in the (m,s) Fock space%

\begin{align}
\psi^{\lbrack A,\dot{B}]}(x)  &  =\frac{1}{(2\pi)^{3/2}}\int\{e^{-ipx}%
\sum_{s_{3}}u(p_{1},s_{3})a(p_{1},s_{3})+\label{field}\\
&  +e^{ipx}\sum_{s_{s}}v(p_{1},s_{3})b^{\ast}(p_{1},s_{3})\}\frac{d^{3}%
p}{2\omega}\nonumber
\end{align}
where $a^{\#},b^{\#}$ are creation/annihilation opertors of Wigner (m,s)
particles and $\psi^{\lbrack A,\dot{B}]}$ are covariant dotted/undotted fields
in the SL(2,C) spinor formalism, it is only necessary to find intertwiners%

\begin{equation}
u(p)D^{(s)}(\tilde{R}(\tilde{\Lambda},p))=D^{[A,\dot{B}]}(\tilde{\Lambda
})u(\tilde{\Lambda}^{-1}p) \label{1}%
\end{equation}
between the Wigner $D^{(s)}(\tilde{R}(\tilde{\Lambda},p))$ and the covariant
$D^{[A,\dot{B}]}(\tilde{\Lambda})$ and these exist for all $A,\dot{B}$ which
relative to the given s obey
\begin{equation}
\mid A-\dot{B}\mid\leq s\leq A+\dot{B} \label{2}%
\end{equation}
For each of these infinitely many values $(A,\dot{B})$ there exists a rectangular

($2A+1)(2\dot{B}+1)\times(2s+1)$ intertwining matrix $u(p).$ Its explicit
construction using Clebsch-Gordan methods can be found in Weinberg's book
\cite{Wei}. Analogously there exist antiparticle (opposite charge)
intertwiners $v(p)$: $D^{(s)\ast}(R(\Lambda,p)\longrightarrow D^{[A,\dot{B}%
]}(\Lambda)$. All of these mathematically different fields in the same Fock
space describe the same physical reality; they are just the linear part of a
huge local equivalence class and they do not exhaust the full ``Borchers
class'' which consists of all Wick-ordered polynomials of the $\psi^{\lbrack
A,\dot{B}]}.$ They generate the same net of local operator algebras and in
turn furnish the singular coordinatizations. Free fields for which the full
content of formula (\ref{field}) can be described by the totality of all
solutions of an Euler-Lagrange equation exist for each (m,s) but are very rare
(example Rarita-Schwinger for s=$\frac{3}{2}$). It is a misconception that
they are needed for physical reason. The causal perturbation theory can be
done in any of those field coordinates and that one needs Euler-Lagrange
fields in the setting of Euclidean functional integrals is an indication that
differential geometric requirements and quantum physical ones do not always go
into the same direction.

On the other hand our modular method for the construction of localized spaces
and algebras use only the minimal intertwiners which are described by square
$\left(  2s+1\right)  \times\left(  2s+1\right)  $ matrices. Without their use
there would be no purely analytic characterization of the domain of the
modular Tomita S-operator.

\section{Vacuum polarization and breakdown of functorial relations}

The functorial relation of the previous section between Wigner subspaces and
operator algebras are strictly limited to the standard halfinteger spin
representations for which generating pointlike free fields exist. The
noncompactly localizable exceptional Wigner representations (anyonic spin,
faithful spin-tower representations of the massless little group) as well as
interacting theories involving standard (halfinteger spin/helicity) particles
do not permit a direct functorial relations between wave function spaces and
operator algebras.

In order to understand the physical mechanism which prevents a functorial
relation it is instructive to look directly to the operators algebras. Given
an operator algebra $\mathcal{A}(\mathcal{O})$ localized in a causally closed
region $\mathcal{O}$ with a nontrivial causal complement $\mathcal{O}^{\prime
}$ (so that ($\mathcal{A}(\mathcal{O}),\Omega$) is standard pair) we may ask
whether this algebra admits a ``polarization-free-generator'' (PFG) namely an
affiliated possibly unbounded closed operator $G$ such that $\Omega$ is in the
domain of $G,G^{\ast}$ and $G\Omega$ and $G^{\ast}\Omega$ are vectors in
$E_{m}H$ with $E_{m}$ projector on the one-particle space.

It turns out that if one admits very crude localizations as that in wedges
then one can reconcile the standardness of the pair $(\mathcal{A}(W),\Omega)$
(i.e. physically the unique $A\Omega\leftrightarrow A\in\mathcal{A}(W)$
relationship) with the absense of polarization clouds caused by localization.
For convenience of the reader we recall the abstract theorem from modular
theory whose adaptation to the local quantum physical situation at hand will
supply the existence of wedge-affiliated PFGs.

An interesting situation emerges if these operators which always generate a
dense one-particle subspace also generate an algebra of unbounded operators
which is affiliated to a corresponding von Neumann algebra $\mathcal{A}%
(\mathcal{O}).$ For causally complete sub-wedge regions $\mathcal{O}$ such a
situation inevitably leads to interaction-free theories i.e. the local
algebras generated by ordinary free fields are the only $\mathcal{A}%
(\mathcal{O})$-affiliated PFGs. Such a situation is achieved by domain
restrictions on the (generally unbounded) PFGs. Without any further domain
restriction on these (generally unbounded) operators it would be difficult to
imagine a constructive use of PFGs.

Before studying PFGs it is helpful to remind the reader of the following
theorem of general modular theory.

\begin{theorem}
Let S be the modular operator of a general standard pair $(\mathcal{A}%
,\Omega)$ and let $\Phi$ be a vector in the domain of S. There exists a unique
closed operator F affiliated with $F$ (notation $F\eta\mathcal{A)}$ which
together with F$^{\ast}$ has the reference state $\Omega$ in its domain and
satisfies
\begin{equation}
F\Omega=\Phi,\,\,F^{\ast}\Omega=S\Phi\label{abstract}%
\end{equation}
\end{theorem}

A proof of this and the following theorem can be found in \cite{BBS}.

For the special field theoretic case $(\mathcal{A}(W),\Omega),$ the domain of
$S$ which agrees with that of $\Delta^{\frac{1}{2}}=e^{\pi K},K=$ boost
generator has evidently a dense intersection $\mathcal{D}^{(1)}=H^{(1)}%
\cap\mathcal{D}_{\Delta^{\frac{1}{2}}}$ with the one-particle space
$H^{(1)}=E_{m}H.$ Hence the operator $F$ for $\Phi^{(1)}\in\mathcal{D}^{(1)}$
is a PFG $G$ as previously defined. However the abstract theorem contains no
information on whether the domain properties admit a repeated use of PFGs
similar to smeared fields in the Wightman setting, nor does it provide any
clew about the position of a $domG$ relative to scattering states. Without
such a physically motivated input, wedge-supported PFGs would not be useful.
An interesting situation is encountered if one requires the $G$ to be
tempered. Intuitively speaking this means that $G(x)=U(x)GU(x)^{\ast}$ has a
Fourier transform as needed if one wants to use PFGs in scattering theory. If
one in addition assumes that the wedge algebras to which the PFGs are
affiliated are of the are of the standard Bose/Fermi type i.e. $\mathcal{A}%
(W^{\prime})=\mathcal{A}(W)^{\prime}$ or the twisted Fermi commutant
$\mathcal{A}(W)^{tw}$, one finds

\begin{theorem}
PFGs for the wedge localization always region exist, but the assumption that
they are tempered leads to a purely elastic scattering matrix $S_{scat}%
=S_{el},$ whereas in d%
$>$%
1+1 is only consistent with $S_{scat}=1$.
\end{theorem}

Together with the recently obtained statement about the uniqueness of the
inverse problem in the modular setting of AQFT \cite{S-matrix} one finally
arrives at the interaction-free nature in the technical sense that the PFGs
can be described in terms of free Bose/Fermi fields.

The nonexistence of PFGs in interacting theories for causally completed
localization regions smaller than wedges (i.e. intersections of two or more
wedges) can be proven directly i.e. without invoking scattering theory

\begin{theorem}
PFGs localized in smaller than wedge regions are (smeared) free fields. The
presence of interactions requires the presence of vacuum polarization in all
state vectors created by applying operators affiliated with causally closed
smaller wedge regions.
\end{theorem}

The proof of this theorem is an extension of the ancient theorem \cite{St-Wi}
that pointlike covariant fields which permit a frequency decomposition (with
the negative frequency part annihilating the vacuum) and commute/anticommute
for spacelike distances are necessarily free fields in the standard sense. The
frequency decomposition structure follows from the PFG assumption and the fact
that in a given wedge one can find PFGs whose localization is spacelike
disjoint is sufficient for the analytic part of the argument to still go
through, i.e. the pointlike nature in the old proof is not necessary to show
that the (anti)commutator of two spacelike disjoint localized PFGs is a
c-number (which only deviates from the Pauli-Jordan commutator by its lack of
covariance). The most interesting aspect of this theorem is the inexorable
relation between interactions and the presence of vacuum polarization which
for the first time leads to a completely intrinsic definition of interactions
which is not based on the use of Lagrangians and particular field coordinates.
This poses the interesting question how the shape of localization region (e.g.
size of double cone) and the type of interaction is related with the form of
the vacuum polarization clouds which necessarily accompany a one-particle
state. We will have some comments in the next section.

As Mund has recently shown, this theorem has an interesting extension to d=1+2
QFT with braid group (anyon) statistics.

\begin{theorem}
(\cite{M}) There are no PFGs affiliated to field algebras localized in
spacelike cones with anyonic commutation relations i.e. sub-wedge localized
fields obeying braid group commutation relations applied to the vacuum are
always accompanied by vacuum polarization clouds. Even in the absence of any
genuine interactions this vacuum polarization is necessary to sustain the
braid group statistics and maintain the spin-statistics relation.
\end{theorem}

This poses the interesting question whether quantum mechanics is compatible
with a nonrelativistic limit of braid group statistics. The nonexistence of
vacuum polarization-free locally (sub-wedge) generated one particle states
suggests that as long as one maintains the spin-statistics connection
throughout the nonrelativistic limit procedure, the result will preserve the
vacuum polarization contributions and hence one will end up with
nonrelativistic field theory instead of quantum mechanics\footnote{The
Leinaas-Myrheim geometrical arguments \cite{L-M} do not take into account the
true spin-statistics connection.}.

Using the concept of PFGs one can also formulate this limitation of quantum
mechanics in a more provocative way by saying that (using the generally
accepted fact that QFT is more fundamental than QM) QM owes its physical
relevance to the fact that the permutation group (Boson/Fermion) statistics
permits sub-wedge localized PFGs (free fields which create one particle states
without vacuum polarization admixture) whereas the more general braidgroup
statistics does not.

Another problem which even in the Wigner setting of noninteracting particles
is interesting and has not yet been fully understood is the pre-modular theory
for disconnected or topologically nontrivial regions e.g. in the simplest case
for disjoint double intervals of the massless $s=\frac{1}{2}$ chiral model on
the circle. Such situations give rise to nongeometric (fuzzy) ``quantum
symmetries'' of purely modular origin without a classical counterpart.

\section{Construction of models via modular localization}

Since up to date more work had been done on the modular construction of d=1+1
factorizing models, we will first illustrate our strategy in that case and
then make some comments of how we expect our approach to work in the case of
higher dimensional d=1+2 anyons and $d\geq1+3$ spin towers.

The construction consists basically of two steps, first one classifies the
possible algebraic structures of tempered wedge-localized PFGs and then one
computes the vacuum polarization clouds of the operators belonging to the
double cone intersections.

Let us confine ourself to the simplest model which we may associate with a
massive selfconjugate scalar particle. If there would be no interactions the
appropriate theorem of the previous section would only leave the free field
which is a PFG for any localization%
\begin{align}
A(x)  &  =\frac{1}{\sqrt{2\pi}}\int\left(  e^{-ip(\theta)x}a(\theta
)+e^{ip(\theta)x}a^{\ast}(\theta)\right)  d\theta\\
A(f)  &  =\int A(x)\hat{f}(x)d^{2}x=\frac{1}{\sqrt{2\pi}}\int_{C}%
a(\theta)f(\theta)d\theta,\text{ }supp\hat{f}\in W\nonumber\\
p(\theta)  &  =m(\cosh\theta,\sinh\theta)\nonumber
\end{align}
where in order to put into evidence that the mass shell only carries one
parameter, we have used the rapidity parametrization in which the plane wave
factor is an entire function in the complex extension of $\theta$ with
$p(\theta-i\pi)=-p(\theta).$ The last formula for the smeared field with the
localization in the right wedge has been written to introduce a useful
notation; the integral extends over the upper and lower conture $C:\theta$ and
$\theta-i\pi,-\infty<\theta<\infty$ where the Fourier transform $f(\theta)$ is
analytic and integrable in the strip which $C$ encloses as a result of its
x-space test function support property. Knowing that tempered PFGs only permit
elastic scattering (see previous section), we make the ``nonlocal'' Ansatz%

\begin{align}
G(x)  &  =\frac{1}{\sqrt{2\pi}}\int\left(  e^{-ipx}Z(\theta)+e^{ipx}Z^{\ast
}(\theta)\right)  d\theta\label{PFG}\\
G(\tilde{f})  &  =\frac{1}{\sqrt{2\pi}}\int_{C}Z(\theta)f(\theta
)d\theta\nonumber
\end{align}
where the $Zs$ are defined on the incoming n-particle vectors by the following
formula for the action of $Z^{\ast}(\theta)$ for the rapidity-ordering
$\theta_{i}>\theta>\theta_{i+1},\,\,\theta_{1}>\theta_{2}>...>\theta_{n}$
\begin{align}
&  Z^{\ast}(\theta)a^{\ast}(\theta_{1})...a^{\ast}(\theta_{i})...a^{\ast
}(\theta_{n})\Omega=\label{bound}\\
&  S(\theta-\theta_{1})...S(\theta-\theta_{i})a^{\ast}(\theta_{1})...a^{\ast
}(\theta_{i})a^{\ast}(\theta)...a^{\ast}(\theta_{n})\Omega\nonumber\\
&  +contr.\,from\text{ }bound\,states\nonumber
\end{align}
In the absence of bound states (which we assume in the following) this amounts
to the commutation relations\footnote{In the presence of bound states such
commutation relations only hold after applying suitable projection operators.}%

\begin{align}
Z^{\ast}(\theta)Z^{\ast}(\theta^{\prime})  &  =S(\theta-\theta^{\prime
})Z^{\ast}(\theta^{\prime})Z^{\ast}(\theta),\,\theta<\theta^{\prime}%
\label{ab}\\
Z(\theta)Z^{\ast}(\theta^{\prime})  &  =S(\theta^{\prime}-\theta)Z^{\ast
}(\theta^{\prime})Z(\theta)+\delta(\theta-\theta^{\prime})\nonumber
\end{align}
where the structure functions $S$ must be unitary in order that the
$Z$-algebra be a $^{\ast}$-algebra. It is easy to show that the domains of the
$Zs$ are identical to free field domains. We still have to show that our
``nonlocal'' $Gs$ are wedge localized. According to modular theory for this we
have to show the validity of the KMS condition. It is very gratifying that the
KMS condition for the requirement that the $G(\tilde{f})$ $supp\tilde
{f}\subset W$ are affiliated with the algebra $\mathcal{A}(W)$ is equivalent
with the crossing property of the $S.$

\begin{proposition}
The PFG's with the above algebraic structure for the Z's are wedge-localized
if and only if the structure coefficients $S(\theta)$ in (\ref{ab}) are
meromorphic functions which fulfill crossing symmetry in the physical $\theta
$-strip i.e. the requirement of wedge localization converts the Z-algebra into
a Zamolodchikov-Faddeev algebra.
\end{proposition}

Improving the support of the wedge-localized test function in $G(\hat{f})$ by
choosing the support of $\hat{f}$ in a double cone well inside the wedge does
not improve $locG(\hat{f}),$ it is still spread over the entire wedge. This is
similar to the spreading property of (\ref{spread}) and certainly very
different from the behavior of smeared pointlike fields.

By forming an intersection of two oppositely oriented wedge algebras one can
compute the double cone algebra or rather (since the control of operator
domains has not yet been accomplished) the spaces of double-cone localized
bilinear forms (form factors of would be operators).

The most general operator $A$ in $\mathcal{A}(W)$ is a LSZ-type power series
in the Wick-ordered $Zs$%

\begin{align}
A  &  =\sum\frac{1}{n!}\int_{C}...\int_{C}a_{n}(\theta_{1},...\theta
_{n}):Z(\theta_{1})...Z(\theta_{n}):d\theta_{1}...d\theta_{n}\label{series}\\
A  &  \in\mathcal{A}_{bil}(W) \label{bil}%
\end{align}
with strip-analytic coefficient functions $a_{n}$ which are related to the
matrix elements of $A$ between incoming ket and outgoing bra multiparticle
state vectors (formfactors). The integration path $C$ consists of the real
axis (associated with annihilation operators and the line $Im\theta=-i\pi.$
Writing such power series without paying attention to domains of operators
means that we are only dealing with these objects (as in the LSZ formalism) as
bilinear forms (\ref{bil}) or formfactors whose operator status still has to
be settled.

Now we come to the second step of our algebraic construction, the computation
of double cone algebras. The space of bilinear forms which have their
localization in double cones are characterized by their relative commutance
(this formulation has to be changed for Fermions or more general objects) with
shifted generators $A^{(a)}(f)\equiv U(a)A(f)U^{\ast}(a)$%
\begin{align}
\left[  A,A^{(a)}(f)\right]   &  =0,\,\forall f\,\,suppf\subset
W\,\label{inter}\\
A  &  \subset\mathcal{A}_{bil}(C_{a})\nonumber
\end{align}
where the subscript indicates that we are dealing with spaces of bilinear
forms (formfactors of would-be operators localized in $C_{a}$) and not yet
with unbounded operators and their affiliated von Neumann algebras. This
relative commutant relation \cite{JMP} on the level of bilinear forms is
nothing but the famous ''kinematical pole relations'' which relate the even
$a_{n}$ to the residuum of a certain pole in the $a_{n+2}$ meromorphic
functions. The structure of these equations is the same as that for the
formfactors of pointlike fields; but whereas the latter lead (after splitting
off common factors \cite{Kar} which are independent of the chosen field in the
same superselection sector) to polynomial expressions with a hard to control
asymptotic behavior, the $a_{n}$ of the double cone localized bilinear forms
are solutions which have better asymptotic behavior controlled by the
Pailey-Wiener-Schwartz theorem. We will not discuss here the problem of how
this improvement can be used in order to convert the bilinear forms into
genuine operators. Although we think that this is largely a technical problem
which does not require new concepts, the operator control of the second step
is of course important in order to convince our constructivist friends that
modular methods really do provide a rich family of nontrivial d=1+1 models. We
hope to be able to say more in future work.

The extension to the general factorizing d=1+1 models should be obvious. One
introduces multi-component $Zs$ with matrix-valued structure functions
$S.\,$\ The contour deformation from the original integral to the ``crossed''
contour which is necessary to establish the KMS conditions in the presence of
boundstate poles in the physical $\theta$-strip compensates those pole
contributions against the boundstate contributions in the state vector Ansatz
(\ref{bound}) \cite{JMP}. The fact that the structure matrix $S(\theta
-\theta^{\prime})$ is the 2-particle matrix element of the elastic S-matrix of
the constructed algebraic net of double cone algebras is not used in this
construction. Of the two aspects of an S-matrix in local quantum physics
namely the large time LSZ (or Haag-Ruelle) scattering aspect and that of the
S-matrix as a relative modular invariant of the wedge algebra we only utilized
the latter.

As a side remark we add that the $Z^{\#}$ operators are conceptually somewhere
between the free incoming and the interacting Heisenberg operators in the
following sense: whereas any particle state in the theory contributes to the
structure of the Fock space and has its own incoming creation/annihilation
operator, the $Z^{\#}$ operators are (despite the rather rough wedge
localization properties of their spacetime related PFGs $G$) similar to
charge-carrying local Heisenberg operators in the sense that all other
operators belonging to particles whose charge is obtained by fusing that of
$Z$ and $Z^{\ast}$ are functions of $Z$ \cite{Zi}$.$ The particle-field
duality which holds for free fields becomes already incalidated by the
interacting wedge-localized PFG $G$ before one gets to the
double-cone-localized operators.

Let us finally make some qualitative remarks about a possible adaptation of
the above two-step processs to the higher dimensional exceptional Wigner
cases. Since their are many wedges, one uses a $\theta$-ordering with respect
to the standard wedge as in \cite{BBS}. Then the nongeometrical nature of the
twist modification $\frak{t}$ of the spatial $\frak{j}$ operator in the Wigner
representation leads to a field-theoretic twist operator $T$ which is the
analog of the $S_{el}$ operator in the previous discussion. This $T$ is
responsible for the modification similar to (\ref{ab}), but this time with
piecewise constant structure constants in the $Z$-analogs which still refer to
the standard wedge ($R$-operators acting on the tower indices in case of spin
towers). With other words the wedge formalism with respect to the standard
wedge is like a tensor product formalism i.e. the n-``particle'' states are
analog to n-fold tensor products in a Fock space. The mismatch between the
algebraic commutant and the geometric opposite of the wedge algebra is
responsible for a drastic modification of the Bisognano-Wichmann theorem and
leads to braid commutation relations between wedge and opposite wedge
operators. The next step namely the formation of the intersection is analog to
the previous case except that instead of a lightlike translation we now have
to take the orthogonal wedge intersection as in section 2.2. The intersection
naturally has to be taken with respect to the twisted relative commutant. It
is expected to build up a rich vacuum polarization structure for the d=1+2
massive anyons as well as for the spin towers.

The impossibility of a compact localization in the case of the exceptional
Wigner representation places them out of reach by Lagrangian quantization
methods. The charge-carrying PFG operators corresponding to the
wedge-localized subspaces as well as their best localized intersections are
more ``noncommutative'' than those for standard QFT and the worsening of the
best possible localization is inexorably interwoven with the increasing
spacelike noncommutativity. This kind of noncommutativity should however be
kept apart from the noncommutativity of spacetime itself whose consistency
with the Wigner representation theory will be briefly mentioned in the
subsequent last section.

\section{Outlook}

In the past the power of Wigner's representation theory has been somewhat
underestimated. As a completely intrinsic relativistic quantum theory which
stands on its own feet (i.e. it does not depend on any classical quantization
parallelism and thus gives quantum theory its deserved dominating position) it
was used in order to back up the Lagrangian quantization procedure \cite{Wei},
but thanks to its modular localization structure it is capable to do much more
and shed new light also on problems which remained outside Lagrangian
quantization and perturbation theory. This includes problems where, contrary
to free fields, no PFG operator (one which creates a pure one-particle state
without a vacuum polarization admixture) for sub-wedge regions exist, but
where wedge-localized algebras still have tempered generators as d=1+1
factorizing models d=1+2 ``free'' anyons and ``free'' Wigner spin towers. It
should however be mentioned that the braid group statistics particles refered
to as anyons associated to d=1+2 continuous spin Wigner representations in
this particular way (i.e. by extending the one-particle twist to multiparticle
states with abelian phase composition) do not exhaust all possibilities of
plektonic statistcs.

Since conformal theories in any dimensions (even beyond chiral theories) are
``almost free'' (in the sense that the only structure which distinguishes them
from free massless theories is the spectrum of anomalous dimension which is
related to an algebraic braid-like structure in timelike direction
\cite{braid}), we believe that they also can be classified and constructed by
modular methods.

This leaves the question of how to deal with interacting massive theories
which have in addition to vacuum polarization real (on shell) particle
creation. For such models PFG generators of wedge algebras are (as a result of
their non-temperedness) too singular objects. One either must hope to find
different (non-PFG) generators, or use other modular methods \cite{Hor}
related to holographically defined modular inclusions or modular
intersections. For example holographic lightfront methods are based on the
observation that the full content of a d-dimensional QFT can be encoded into
d-1 copies of one abstract chiral theory whose relative placement in the
Hilbert space of the d-dimensional theory carries the information. What
remains to be done is to characterize the kind of chiral theory and its
relative positions in a constructively manageable way.

Another insufficiently understood problem is the physical significance of the
infinitely many modular symmetry groups which (beyond the Poincar\'{e} or
conformal symmetry groups which leave the vacuum invariant) act in a fuzzy way
within the localization regions and in their causal complements \cite{fuzzy}.
An educated guess would be that they are related to the nature of the vacuum
polarization clouds which local operators in that region generate from the vacuum.

It is an interesting (and in recent years again fashionable) question whether
besides the macro-causal relativistic quantum mechanics mentioned in the
introduction and the micro-causal local quantum physics there are other
relativistic non-micro causal quantum theories\footnote{A recent paper by Lieb
and Loss \cite{Lieb} contains an interesting attempt to combine relativistic
QM with local quantum field theory. To make this model fully cluster separable
(macro-causal) one probably has to combine the localization properties of
relativistic quantum mechanics with those of modular localization for the
photon field.} which permit at least the physical notion of scattering and
which unlike the the relativistic mechanics preserve some of the vacuum
polarization properties especially those which are necessary to keep the TCP
theorem (so that the existence of antiparticles is an inexorably consequence)
address the question of localization (string theory presently does not; if the
localization discussed there would have the fundamental quantum significance
as the one used in this paper then string theory would be a special kind of
AQFT). All attempts to obtain ultraviolet improved renormalizable theories
naturally after the discovery of renormalization) by allowing nonlocal
interactions, starting from the Kristensen-Moeller-Bloch \cite{M-K}%
\cite{Bloch} replacement of pointlike Lagrangian interactions by formfactors
and the Lee-Wick complex pole modification \cite{L-W} of Feynman rules up to
some of the recent noncommutative spacetime failed. Even if Lorentz invariance
and unitarity (including the optical theorem) could have been maintained in
those proposals, the main reason for original motivation namely ultraviolet
convergence was not borne out \cite{Bloch}. Of course even without this
motivation it would be very interesting to know if there are any ``physically
viable'' nonlocal relatistic theories at all. By this we mean the survival of
the physically indispensible macro-causality\footnote{In case of formfactor
modifications of pointlike interaction vertices this was shown in \cite{Bloch}
and in case of the Feynman rule modifications by complex poles in
\cite{Swie}.}. For the relativistic particle theory mentioned in the
introduction this macro-causality was insured via the cluster-separability
properties of the S-matrix. more than 50 years of history on this issue has
taught time and again that the naive idea that a mild modification of
pointlike Lagrangian interactions will still retain macro-causality turns out
to be wrong under closer scrutiny. The general message is that the notion of a
mild violation of micro-causality (i.e. maintaining macro-causality) within
the standard framework is an extremely delicate concept \cite{cau}.

In more recent times Doplicher Fredenhagen and Roberts \cite{DFR} discovered a
Bohr-Rosenfeld like argument which uses a quasiclassical interpretation of the
Einstein field equation (coupled with a requirement of absence of
measurement-caused black holes which would trap photons) and leads to
uncertainty relations of spacetime. Although the initiating idea was very
conservative, the authors were nevertheless led to quite drastic conceptual
changes since the localization indexing of field theoretic observables is now
done in terms of noncommutative spacetime in which points correspond to pure
states on a quantum mechanical spacetime substrate on which the Poincar\'{e}
group acts. They found a model which saturate their commutation relations and
they started to study QFTs over this new structure. In more recent times it
was realized \cite{BDF}, that when one recast such models into the setting of
Yang-Feldman perturbation theory with nonlocal interactions, many problems
which appear if one does not rethink the formalism but just copies old
perturbative recipes from the standard case \cite{Filk} (as violation of
L-invariance and unitarity, which have their origin in the fact that in the
new context Feynman $i\varepsilon$ prescription is not the same as
time-ordering) \ disappear and the only conceptual problem which remains is an
appropriate form of macro-causality. Interestingly enough, these were
precisely the techniques used in the first post renormalization investigations
of nonlocal interactions \cite{M-K}.

So there seems to be at least some hope that those specific nonlocalities
caused by those models whose lowest nontrivial perturbative order is discussed
in \cite{BDF} are exempt from the historical lessons. This would be a theory
to which the Wigner approach is applicable and the Fock space structure is
maintained but with different localization concepts. It would be very
interesting indeed if besides the two mentioned relativistic theories build on
different localization concept treated in this article there could exist a
theory of Wigner particles interacting on noncommutative spacetime in a
possibly macro-causal way and uphold the significant gains concerning the TCP
structure and antipartices which are so inexorably linked to vacuum
polarization. Such a quest on a fundamental level should not be confused with
the phenomenological use of the language of noncommutative geometry for
certain conventional Schroedinger systems involving constant magnetic fields
\cite{Douglas} since in those cases the localization concepts of the
Schroedinger theory are in no way affected by the observation that one may
write the system in terms of different dynamical variables.

In this context it is worthwhile to remember that the full local
(anti)commutativity is not used in e.g. the derivation of the TCP theorem
\cite{St-Wi}. Using the present terminology the TCP property is in fact known
to be equivalent to wedge localization. However the question of whether a
modular wedge localization is possible in the context of the correctly
formulated noncommutative L-invariant and unitary models \cite{DFR}\cite{BDF}
\ may well have a positive answer \cite{Fre}. This point is certainly
worthwhile to return to in future work.

It is very regrettable that such conceptually subtle points\footnote{The claim
in \cite{Douglas} that ''noncommutativity of the space-time coordinates
generally conflicts with Lorentz invariance'' contradicts the results of the
1995 seminal paper \cite{DFR} and a fortiori the forthcoming explicit
perturbative model calculations in \cite{BDF}.} seem to go unnoticed in the
new globalized way of doing particle physics \cite{Douglas}. It seems that the
ability of recognizing conceptually relevant points, which has been the
hallmark of part of 20 century physics, has been lost in the semantic efforts
of attaching physical-sounding words to mathematical inventions.

It is well-known to quantum field theorist with some historical awareness that
the role of causality and localization was almost never appreciated/understood
by most mathematicians. This has a long tradition. A good illustration is the
impressive scientific curriculum of Irvine Segal, one of the outstanding
pioneers of the algebraic approach. If in those papers localization concepts
would have been treated with the same depth and care as global mathematical
aspects of AQFT, quantum field theory probably would have undergone a more
rapid development and we would have been spared the many differential
geometric traps and pitfalls, including the banalization of Euclidean methods.

\textbf{Acknowledgements: }One of the authors (B.S.) is indebted to Wolfhardt
Zimmermann for some pleasant exchanges of reminiscences on conceptual problems
of QFT of the 50s and 60s, as well as for related references. B.S. is also
indebted to Sergio Doplicher and Klaus Fredenhagen for an explanation of the
actual status of their 1995 work. Finally the authors would like to thank
Fritz Coester for some valuable email information which influenced the content
of the introduction.

\section{Appendices}

Here we have collected some mathematical details for the convenience of the reader.

\subsection{Appendix A: The abstract spatial modular theory}

Suppose we have a ``standard'' spatial modular situation i.e. a closed real
subspace $H_{R}$ of a complex Hilbert space $H$ such that $H_{R}\cap
iH_{R}=\left\{  0\right\}  $ and the complex space $H_{D}\equiv H_{R}+iH_{R}$
is dense in $H.\,$\ Let $e_{R}$ and $e_{I}$ be the projectors onto $H_{R}$ and
$iH_{R}$ and define operators
\begin{equation}
t_{\pm}\equiv\frac{1}{2}(e_{R}\pm e_{I})
\end{equation}
Because of the reality restriction the two operators have very different
conjugation properties, $t_{+}$ turns out to be positive $0<t_{+}<\mathbf{1}$,
but $t_{-}$ is antilinear. These properties follow by inspection through the
use of the projection- and reality-properties. There are also some easily
derived quadratic relations between involving the projectors and $t\pm$%
\begin{align}
e_{R,I}t_{+}  &  =t_{+}(1-e_{I,R})\\
t_{+}t_{-}  &  =t_{-}(1-t_{+})\nonumber\\
t_{+}^{2}  &  =t_{-}(1-t_{-})\nonumber
\end{align}

\begin{theorem}
(\cite{Rieffel}) In the previous setting there exist modular
objects\footnote{In the physical application the Hilbert space can be
representation space of the Poincar\'{e} group which carries an irreducible
positive energy representation or the bigger Fock space of (free or incoming)
multi-particle states. In order to have a uniform notation we use (different
from section 2) big letters for the modular objects and the transformations,
i.e. $S,J,$ $\Delta,U(a,\Lambda).$} $J$, $\Delta$ and $S=$ $\frak{j}%
\Delta^{\frac{1}{2}}$ which reproduce $H_{R}$ as the +1 eigenvalue real
subspace of $S$. They are related to the previous operators by
\begin{align*}
t_{-}  &  =J\left|  t_{-}\right| \\
\Delta^{it}  &  =\left(  1-t_{+}\right)  ^{it}t_{+}^{-it}%
\end{align*}
\end{theorem}

The proof consists in showing the commutation relation $J\Delta^{it}%
=\Delta^{it}$ $J$ ($\curvearrowright$ $J\Delta=\Delta^{-1}$ $J$ since $J$ is
antiunitary) which establishes the dense involutive nature $S^{2}\subset1$ of
$S$ by using the previous identities. It is not difficult to show that $0$ is
not in the point spectrum of $\Delta^{it}.$

\begin{corollary}
If $H_{R}$ is standard, then $iH_{R},$ $H_{R}^{\perp}$ and $iH_{R}^{\perp}$
are standard. Here the orthogonality $\perp$ refers to the real inner product
$Re(\psi,\varphi).$ Furthermore the $J$ acts on $H_{R}$ as
\[
JH_{R}=iH_{R}^{\perp}%
\]
\end{corollary}

We leave the simple proofs to the reader (or look up the previous reference
\cite{Rieffel}). The orthogonality concept is often expressed in the physics
literature by $iH_{R}^{\perp}=H_{R}^{symp\perp}$ referring to symplectic
orthogonality in the sense of $Im(\psi,\varphi).$ There is also a more direct
analytic characterization of $\Delta$ and $J$

\begin{theorem}
(spatial KMS condition) The functions f(t)=$\Delta^{it}\psi,$ $\psi\in H_{R}$
permits an holomorphic continuation f(z) holomorphic in the strip -$\frac
{1}{2}\pi<\operatorname{Im}z<0,$ continuous and bounded on the real axis and
fulfilling $f(t-\frac{1}{2}i)=$ $Jf(t)$ which relates the two boundaries. The
two commuting operators $\Delta^{it}$ and j are uniquely determined by these
analytic properties i.e. H$_{R}$ does not admit different modular objects.
\end{theorem}

Another important concept in the spatial modular theory is ``modular inclusion''

\begin{definition}
(analogous to Wiesbrock) A inclusion of a standard real subspace $K_{R}$ into
a standard space $K_{R}\subset H_{R}$ is called ``modular'' if the modular
unitary $\Delta_{H_{R}}^{it}$ of $H_{R}$ compresses $K_{R}$ for one sign of t
\[
\Delta_{H_{R}}^{it}K_{R}\subset K_{R}\,\,\,\,t<0
\]
If necessary one adds a -sign i.e. if the modular inclusion happens for t%
$>$%
0 one calls it a $-$modular inclusion.
\end{definition}

\begin{theorem}
The modular group of a modular inclusion i.e. $\Delta_{K_{R}}^{it}$ together
with $\Delta_{H_{R}}^{it}$ generate a unitary representation of the
two-parametric affine group of the line.
\end{theorem}

The proof consists in observing that the positive operator $\Delta_{K_{R}%
}-\Delta_{H_{R}}\geq0$ is essentially selfadjoint. Hence we can define the
unitary group
\begin{equation}
U(a)=e^{i\frac{1}{2\pi}a\overline{(\Delta_{K_{R}}-\Delta_{H_{R}})}}%
\end{equation}
The following commutation relation
\begin{align}
\Delta_{H_{R}}^{it}U(a)\Delta_{H_{R}}^{-it}  &  =U(e^{\pm2\pi t}a)\\
J_{H_{R}}U(a)J_{H_{R}}  &  =U(-a)\nonumber
\end{align}
and several other relations between $\Delta_{H_{R}}^{it},\Delta_{K_{R}}^{it},
$ $J_{H_{R}},$ $J_{K_{R}},U(a).$ The above relations are the
Dilation-Translation relations of the 1-dim. affine group. It would be
interesting to generalize this to the modular intersection relation in which
case one expects to generate the SL(2,R) group.

The actual situation in physics is opposite: from group representation theory
of certain noncompact groups $\pi(G)$ one obtains candidates for $\Delta^{it}$
and $J$ from which one passes to $S$ and $H_{R}.$ In the case of the
Poincar\'{e} or conformal group the boosts or proper conformal transformations
in positive energy representations lead to the above situation. The
representations do not have to be irreducible; the representation space of a
full QFT is also in the application range of the spatial modular theory. If
the positive energy representation space is the Fockspace over a one-particle
Wigner space, the existence of the CCR (Weyl) or CAR functor maps the spatial
modular theory into operator-algebraic modular theory of Tomita and Takesaki.
In general such a step is not possible. Connes has given conditions on the
spatial theory which lead to the operator-algebraic theory. They involve the
facial structure of positive cones associated with the space $H_{R}.$ Up to
now it has not been possible to use them for constructions in QFT. The
existing ideas of combining the spatial theory of particles with the
Haag-Kastler framework of spacetime localized operator algebras uses the
following 2 facts

\begin{itemize}
\item  The wedge algebra $\mathcal{A}(W)$ has known modular objects
\begin{align}
\Delta^{it}  &  =U(\Lambda_{W}(-2\pi t))\\
J  &  =S_{scat}J_{0}\nonumber
\end{align}
Whereas the wedge affiliated L-boost (in fact all P$_{+}^{\uparrow}$
transformations) is the same as that of the interacting or free
incoming/outgoing theory, the interaction shows up in those reflections which
involve time inversion as $J.$ In the latter case the scattering operator
$S_{scat}$ intervenes in the relation between the incoming (interaction-free)
$J_{0}$ and its Heisenberg counterpart $J.$ In the case of interaction free
theories the $J_{0}$ contains in addition to the geometric reflection
(basically the TCP) a ``twist'' operator which is particularly simple in the
case of Fermions.

\item  The wedge algebra $\mathcal{A}(W)$ has PFG-generators. In certain cases
these generators have nice (tempered) properties which makes them useful in
explicit constructions. Two such cases (beyond the standard free fields) are
the interacting d=1+1 factorizing models and the free anyonic and Wigner
spin-tower representations in both cases the PFG property is lost (vacuum
polarization is present) for sub-wedge algebras. In the last two Wigner cases
the presence of the twist requires this, only the fermionic twist in the case
of $S_{scat}=1$ is consistent with having PFGs for all localizations.
\end{itemize}

\subsection{Appendix B: Infinite dimensional covariant representations}

In terms of the little group generators relative to the fixed vector $\frac
{1}{2}(1,0,0,1)$ the Pauli-Lubanski operators has the form
\[
W_{\mu}=-\frac{1}{2}\varepsilon_{\mu\nu\sigma\tau}J^{\nu\sigma}P^{\tau}%
=\frac{1}{2}(M_{3},\Pi_{1},\Pi_{2},M_{3})
\]
where $M_{3}$ is the 3-component of the angular momentum and $\Pi_{i}$ are the
two components of the Euclidean translations which together make up the
infinitesimal generators of $\tilde{E}(2).$ An representation of the little
group can be given in any of the Gelfand at al. irreducible representation
spaces of the homogeneous Lorentz group. These consist of homogeneous
functions of two complex variables $\zeta=$($\zeta_{1},\zeta_{2}$) which are
square integrable with respect to the following measure
\begin{align}
d\mu(\zeta)  &  =\frac{1}{4\pi}\left(  \frac{i}{2}\right)  ^{2}d^{2}\zeta
d^{2}\bar{\zeta}\delta(\frac{1}{2}\zeta\textsl{q}\zeta^{\ast}-1),\,\textsl{q=}%
\sigma^{\mu}q_{\mu},\,\,q^{2}=0,q_{0}>0\\
\left(  f,g\right)   &  =\int d\mu(\zeta)\bar{f}(\zeta)\,g(\zeta),\,\,\,f(\rho
e^{i\alpha}\zeta)=\rho^{2(c-1)}e^{2il_{0}\alpha}f(\zeta),\,\,\lambda_{0}%
=0,\pm\frac{1}{2},\pm1,..,c=i\nu,\nonumber
\end{align}
The inner product is independent of the choice of the lightlike vector $q$ if
$c=i\nu$ because the integrand has total homogeneous degree -4 and on
functions $F(\rho\zeta)=\rho^{-4}F(\zeta)$ with this degree the integral is
q-independent. This family of unitary irreducible representations
$\chi=\left[  \lambda_{0},c=i\nu\right]  $ for $-\infty<\nu<\infty$ of SL(2,C)
is called the \textit{principal series} representation. Another such family,
the \textit{supplementary series }$\chi=\left[  \lambda_{0},c\right]
,\,\,-1<c $ $<1$ contains an additional integral operator $K(\zeta,\eta)$%
\begin{align}
\left(  f,g\right)   &  =\int d\mu(\zeta)\bar{f}(\zeta)\,\int K(\zeta
,\eta)g(\eta)\\
K(\zeta,\eta)  &  =N^{-1}\left(  \eta\varepsilon\zeta\right)  ^{-l_{0}%
-c-1}\overline{\left(  \eta\varepsilon\zeta\right)  }^{-l_{0}-c-1}\nonumber
\end{align}
We now define basisvectors in the above representation spaces which carry a
representation of the little group
\begin{align}
\left(  \Pi_{1}^{2}+\Pi_{2}^{2}\right)  f_{\lambda}^{\chi,\rho}(\zeta)  &
=\rho^{2}f_{\lambda}^{\chi,\rho}(\zeta),\,\,M_{3}f_{\lambda}^{\chi,\rho}%
(\zeta)=-\lambda f_{\lambda}^{\chi,\rho}(\zeta)\\
\left(  U(\tilde{E})f_{\lambda}^{\chi,\rho}\right)  (\zeta)  &  =\sum
_{\lambda^{\prime}}f_{\lambda^{\prime}}^{\chi,\rho}(\zeta\tilde{E}%
)d_{\lambda^{\prime},\lambda}(\tilde{E})\nonumber\\
f_{\lambda}^{\chi,\rho}(\zeta)  &  =\left|  \zeta_{2}\right|  ^{2c-2}%
e^{-i\lambda\phi}J_{l_{0}-\lambda}(2\rho\left|  z\right|  )e^{il_{0}\alpha
},\,\,\phi\nonumber
\end{align}

In a similar way, the d=1+2 anyonic representations may be rewritten in terms
of infinite dimensional covariant representations. It has been shown
\cite{Mu-S} that the following family of covariant unitary representations of
$\mathcal{\tilde{P}}_{3}^{\uparrow}$ are useful in the covariant description
of the (m,s) Wigner representation
\begin{align*}
\left(  U(a,(\gamma,\omega))\psi\right)  (p,z)  &  =e^{ipa}\tau_{h,\sigma
}((\gamma,\omega);z)\psi(\Lambda(\gamma,\omega)^{-1}p,(\gamma,\omega
)^{-1}z)\,\,\\
\tau_{h,\sigma}((\gamma,\omega);z)  &  =e^{-i\omega h}\left(  \frac
{1+z\bar{\gamma}}{1+z^{-1}\gamma}\right)  ^{h}\left(  1+z\bar{\gamma}\right)
^{-1-2\sigma}\left(  1+\left|  \gamma\right|  \right)  ^{\frac{1}{2}+\sigma}\\
(\gamma,\omega)\cdot z  &  =e^{-i\omega}\frac{z-\gamma e^{i\omega}}%
{1-z\bar{\gamma}e^{-i\omega}}%
\end{align*}
Here the $\tau$ are Bargmann's principle series representations of
$\widetilde{SL(2,R)}$ acting on the covering of the circle with the circular
coordinate being $z$, $\left|  z\right|  =1.$ The last formula is the action
of the Moebius group on the circle.$\,$The wave functions $\psi(p,z)\,$\ in
this formula are from $L^{2}(p\in H_{m}^{\uparrow},z=e^{i\varphi};\frac
{dp}{2p_{0}},d\varphi)$ and in the range $-\frac{1}{2}<h\leq\frac{1}%
{2},\,\sigma\in iR$ the action is unitary. It has been shown that this
covariant representation can be decomposed into a direct sum of Wigner
representations $(m,s=k-h).$ $k\in\mathbb{Z}$.

\end{document}